\documentclass{article}
\usepackage{graphicx}
\usepackage{amsmath,amssymb}
\usepackage{graphics,color,array,calc,rotating,epsfig,psfrag}
\numberwithin{equation}{section}
\usepackage{cite}
\usepackage{bm}
\usepackage{dcolumn}  
\usepackage{amsfonts}
\usepackage[mathscr]{eucal}
\def\be{\begin{equation}} \def\ee{\end{equation}}
\def\bea{\begin{eqnarray}} \def\eea{\end{eqnarray}}

\newcommand{\ie}{{\it i.e.,}\ }

\newcommand\prt{\partial}

\newcommand{\nn}{\nonumber}
\begin{document}
\baselineskip 18pt%
\begin{titlepage}
\vspace*{1mm}%
\hfill%
\vspace*{15mm}%
\hfill
\vbox{
    \halign{#\hfil         \cr
          } 
      }  
\vspace*{20mm}

\centerline{{\large {\bf Stability of vacua in New Massive Gravity in different gauges}}}
\vspace*{5mm}
\begin{center}
{ Ahmad Ghodsi\footnote{ahmad@ipm.ir} and Davood Mahdavian Yekta \footnote{da.\,mahdavianyekta@stu-mail.um.ac.ir}}\\
\vspace*{0.2cm}
{ Department of Physics, Ferdowsi University of Mashhad, \\
P.O. Box 1436, Mashhad, Iran}\\
\vspace*{0.1cm}
\end{center}

\begin{abstract} 
We consider $AdS_3$ and warped $AdS_3$ vacua in new massive gravity and study the highest weight modes and  general propagating modes as a set of solutions for the linearized equations of motion. We observed that depending on the choice of gauge there are two types of solutions. We show that for warped $AdS_3$ vacuum, the massless modes which appear only  in the harmonic gauge have zero energy density and do not get higher curvature corrections. By computing the energy density it can be shown that all massive modes have negative energy density. Our computations prove that the massive modes in warped $AdS_3$ cannot be excluded by an appropriate boundary condition and this makes the theory unstable.
\end{abstract} 

\end{titlepage}

\section{Introduction}
According to $AdS/CFT$ correspondence \cite{Maldacena:1997re}, each field propagating on $AdS$ space is in a one to one correspondence with an operator in the Conformal Field Theory (CFT) which lives on the boundary of $AdS$. The boundary fields parametrize the boundary conditions of the bulk propagating fields and couple to the operators of the dual CFT. 

The sub-leading radial behavior of the propagating fields at boundary is obtained by finding the most general asymptotic solutions for the field equations. For theories that admit asymptotically locally $AdS$ solutions these general solutions, which are sometimes called the Fefferman-Graham expansion, can always be found by solving algebraic equations\cite{Skenderis:2002wp}.

Appropriate boundary conditions control the bulk propagating fields and this will be important when
there are negative energy propagating modes which make the theory unstable. In this paper, we study this problem in the context of $AdS_3/CFT_2$.

The three-dimensional gravity described by Einstein-Hilbert (EH) action, has no degrees of freedom \cite{Deser:1984tn,Deser:1984dr}, and usually higher order derivative deformations of pure EH gravity provide the theory with propagating degrees of freedom, i.e. three-dimensional gravitons. The first theory of this type was the topologically massive gravity (TMG) which was constructed by adding a cosmological constant and a gravitational Chern-Simons term\cite{Deser:1982vy} \!-\! \cite{Banados:1992wn}. Another theory which we are going to consider here, is the New Massive Gravity (NMG)\cite{Bergshoeff:2009hq}. Also some extended theories of NMG have been discussed in \cite{Gullu:2010pc}, \cite{Ghodsi:2011ua},\cite{Ghodsi:2010ev},\cite{Sinha:2010ai}. The quantization of these theories seems to give a richer structure than the EH theory and provides interesting toy models for higher-dimensional theories of quantum gravity.

$AdS_3$ vacuum is the first known solution of the massive gravity models. Another vacuum solution of the higher order derivative actions is Warped-$AdS_3$ ($WAdS_3$). These vacua admit black hole solutions known as BTZ \cite{Banados:1992wn} and warped-$AdS_3$ black holes\cite{Anninos:2008fx,Clement:2009gq}. In this paper we are going to discuss the asymptotic behavior of the metric fluctuations in NMG model. Since the corresponding black holes have equivalent asymptotic behavior we only consider  the $AdS_3$ and $WAdS_3$ vacua here.

To find the behavior of metric fluctuations around a background there are two main approaches. In the first approach, one finds the highest weight modes corresponding to representations of the isometry group of the background. Knowing that one can associate generators to the isometry group of a background ($SL(2,R)_L\times SL(2,R)_R$ for $AdS_3$ or $SL(2,R)\times U(1)$ for $WAdS_3$) the highest weight modes are defined as those modes which are annihilated  by raising operators ($L_1$ and $\bar{L}_1$ generators in $AdS_3$ or $L_1$ generator in $WAdS_3$). Moreover, these modes must be solutions of the linearized equations of motion.  

In the second approach, one substitutes the metric fluctuations as  eigen-modes of energy and momentum in the linearized equations of motion and  tries to find a decoupled differential equation for each component of the perturbations. A regular solution at the asymptotic limit can be obtained by using the Frobenius's method. In the asymptotic limit where the radial direction $r$, approaches to the boundaries, solutions behave as $r^{-B}$, where $B$ depends on the parameters of the theory. It can be shown that $B$ is  closely related to the frequencies in the highest weight approach. These asymptotic solutions are often called the general propagating modes. 
For example in TMG the calculations for $WAdS_3$ vacuum has been discussed in \cite{Anninos:2009zi} by these two approaches. 

The stability around a certain background depends on the selection of consistent boundary conditions. For example there exist several consistent choices of boundary conditions for three dimensional massive gravity models in $AdS_3$. But only a special class of these boundary conditions exclude the negative energy modes (unstable modes), which are distinct from the Brown\,-\,Henneaux (BH) boundary conditions\cite{BH}. 

Similarly the massive propagating modes of $WAdS_3$ do not obey the Comp\'{e}re\,-\,Detournay (CD) boundary conditions \cite{Compere:2007in,Compere:2008cv}. In fact the BH and CD boundary conditions are only consistent with the pure large gauge in which the asymptotic perturbations are given by the Lie derivative of the background metric $h_{\mu\nu}={\mathcal L}_{\xi}\bar g_{\mu\nu}$ where $\xi^{\mu}$ is the non-vanishing asymptotic diffeomorphism. We have been considered these boundary conditions in \cite{Ghodsi:2011ua} which lead to the central charges of the dual CFT living at the boundary of asymptotically $AdS_3$ space-times. 

The stability of TMG around $AdS_3$ vacuum in harmonic gauge has been shown in \cite{Li:2008dq}. They have obtained a stable theory without negative energy at chiral point $\mu l=1$. Calculations for stability of $AdS_3$ and BTZ black holes in NMG have been done in\cite{Liu:2009bk}, \cite{Myung:2011bn} and \cite{Afshar:2010ii}.

In this paper we will study the metric perturbations around the $AdS_3$ and $WAdS_3$ backgrounds in NMG in two different gauge conditions, the  harmonic (transverse) gauge $\nabla_{\mu}\,h^{\mu\nu}=0$ and $h_{\mu\varphi}=0$ gauge. We will check the stability of all modes explicitly.

Even though higher derivative terms are treated as perturbative corrections to two derivative Lagrangians, they do not change the usual $AdS/CFT$ setup\cite{Skenderis:2009nt}. We will consider the effect of these terms on the spectrum of the theory.

This paper is organized as follows: In section 2 we briefly discuss the Lagrangian of NMG and its equations of motion and extract the linearized form of these equations around an arbitrary background. In section 3 we review and study $AdS_3$ vacuum and its perturbations in different gauges and then check the stability of this vacuum. In section 4 we repeat all steps in section three but for $WAdS_3$ vacuum. In section 5 we discuss the stability conditions for warped solution in different gauges. We compute the energy density for all possible modes. In section 6 we will consider extended NMG model, which contains up to sixth order derivative terms and study the modifications of different modes. Section 7 includes summary and discussions.

\section{NMG and its linearized equations of motion}
The new massive gravity is given by the following Lagrangian \cite{Bergshoeff:2009hq} 
\bea 
\label{lagnmg} \mathcal{L}=\sqrt{-g}\,\, \big(R-2\Lambda+\kappa_1\,R^2+\kappa_2\,R_{\mu\nu}R^{\mu\nu}\big)\,,
\eea 
where  $\Lambda$ is the cosmological constant. Together with a three dimensional gravitational coupling constant, there are two other couplings, $\kappa_1=\frac{3}{8m^2}$ and $\kappa_2=-\frac{1}{m^2}$, where $m$ is a mass parameter.
The equations of motion  are given by
\bea\label{eomnmg} 
T^{NMG}_{\mu\nu}&=&R_{\mu\nu}-\frac12 g_{\mu\nu} R+\Lambda g_{\mu\nu}+2\kappa_1  R (R_{\mu\nu}-\frac14 g_{\mu\nu} R)+(2\kappa_1+\kappa_2) (g_{\mu\nu}\Box-\nabla_{\mu}\nabla_{\nu})R\nn\\
&+&\kappa_2  \Box ( R_{\mu\nu}-\frac12g_{\mu\nu} R)+2\kappa_2 (R_{\mu\rho\nu\sigma}-\frac14 g_{\mu\nu} R_{\rho\sigma}) R^{\rho\sigma}\,,
\eea
which correspond to variation of the gravitational fields.

To study behavior of the gravitational fluctuations, we use the linearized  equations of motion around an arbitrary background. Let $\bar g_{\mu\nu}$ be a background metric and its fluctuation is given by $h_{\mu\nu}$, so the total geometry is describing by $g_{\mu\nu}\equiv\bar{g}_{\mu\nu}+h_{\mu\nu}$. Consequently the Christoffel connection,  Riemann and  Ricci tensors are linearized  as follows
\bea 
\delta \Gamma^\mu_{\rho\sigma}&=&\frac12(\bar{\nabla}_{\rho}h^{\mu}_{\sigma}+\bar{\nabla}_{\sigma}h^{\mu}_{\rho}-\bar{\nabla}^{\mu}h_{\rho\sigma})\,,\quad
\delta {R^{\mu}}_{\rho\nu\sigma}=\bar{\nabla}_{\nu}\,\delta \Gamma^{\mu}_{\rho\sigma}-\bar{\nabla}_{\sigma}\,\delta\,\Gamma^{\mu}_{\rho\nu}\,,\nn\\
\delta R_{\mu\nu}&=&\frac12\,\big(\bar{\nabla}_{\alpha}\bar{\nabla}_{\mu}h^{\alpha}_{\nu}+\bar{\nabla}_{\alpha}\bar{\nabla}_{\nu}h^{\alpha}_{\mu}-\bar{\nabla}_{\mu}\bar{\nabla}_{\nu}h-\bar{\nabla}_{\alpha}\bar{\nabla}^{\alpha}h_{\mu\nu}\big)\,,\quad
\delta R=\bar g^{\alpha\beta}\,\delta R_{\alpha\beta}-h^{\alpha\beta}\,\bar R_{\alpha\beta}\,.
\eea

All derivatives are taken with respect to the background metric $\bar g_{\mu\nu}$. Employing the above relations, the linearized energy-momentum tensor for new massive gravity becomes
\bea \label{leomnmg}
\delta T^{NMG}_{\mu\nu}&=&\delta R_{\mu\nu}-\frac12 h_{\mu\nu}(\bar R-2\Lambda)-\frac12 \bar g_{\mu\nu}\, \delta R+2\kappa_1 (\bar R_{\mu\nu} \, \delta R+\bar{R} \,\delta R_{\mu\nu}-\frac14 h_{\mu\nu}\, \bar {R}^2-\frac12  \,\bar g_{\mu\nu} \,\bar{R} \,\delta R)\nn\\
&+&(2\kappa_1+\kappa_2)\big(\bar g_{\mu\nu}\bar{\Box}-\bar{\nabla}_{\mu}\bar{\nabla}_{\nu}\big)\delta R+\kappa_2 \big(\bar{\Box}\delta R_{\mu\nu}-\frac12 \bar g_{\mu\nu} \bar{\Box} \delta R-h^{\alpha\beta} \bar{\nabla}_{\alpha}\bar{\nabla}_{\beta}\bar R_{\mu\nu}-\bar g^{\alpha\beta}\delta \Gamma^{\lambda}_{\alpha\beta} \bar{\nabla}_{\lambda}\bar R_{\mu\nu}\nn\\
&-&\delta \Gamma^{\lambda}_{\alpha\mu} \bar{\nabla}^{\alpha}\bar R_{\lambda\nu}-\delta \Gamma^{\lambda}_{\alpha\nu} \bar{\nabla}^{\alpha}\bar R_{\mu\lambda}-\bar{\nabla}^{\alpha}(\delta \Gamma^{\lambda}_{\beta\mu} \bar R_{\lambda\nu}+\delta \Gamma^{\lambda}_{\beta\nu} \bar R_{\mu\lambda})\big)
+2 \kappa_2\big(\delta R_{\mu\rho\nu\sigma} \bar R^{\rho\sigma}+\bar{R}_{\mu   \nu}^{   \rho   \sigma}  \delta R_{\rho\sigma}\nn\\
&-&2 h^{\rho\lambda} \bar R_{\mu\rho\nu\sigma} {{\bar R}_{\lambda}}^{\sigma}-\frac14 h_{\mu\nu} {\bar R_{\rho\sigma}}^2+\frac12 \bar g_{\mu\nu} h^{\rho\lambda} \bar R_{\lambda}^{   \sigma} \bar R_{\rho\sigma}-\frac12 \bar g_{\mu\nu} \bar R^{\rho\sigma} \delta R_{\rho\sigma}\big)\,.
\eea
Everywhere we have used a bar notation, means that a quantity must be computed in the background metric.
\section{$AdS_3$ vacuum in NMG}
The global $AdS_3$ metric is a vacuum solution for NMG equations of motion (\ref{eomnmg}) 
\be\label{adsmetric}
ds^2={l^2}\Big[-(1+r^2) d\tau^2+\frac{dr^2}{1+r^2}+r^2 d\varphi^2\Big]\,,
\ee
where $l$ is the radius of $AdS_3$ space. 
The isometry group for this space is $SL(2,R)_L\times SL(2,R)_R$ with the following  left and right moving set of generators
\bea
&&L_0=\frac{i}{2}(\prt_t+\prt_{\varphi})\,,\quad
L_{\pm 1}=\frac{i}{2}\,e^{\pm i(t+\varphi)}\,\Big[\frac{r}{\sqrt{1+r^2}}\,\prt_{t}\mp i \sqrt{1+r^2}\prt_{r}+\frac{\sqrt{1+r^2}}{r}\prt_{\varphi} \Big]\,,\nn\\
&&\bar L_0=\frac{i}{2}(\prt_t-\prt_{\varphi})\,,\quad
\bar L_{\pm 1}=\frac{i}{2}\,e^{\pm i(t-\varphi)}\,\Big[\frac{r}{\sqrt{1+r^2}}\,\prt_{t}\mp i \sqrt{1+r^2}\prt_{r}-\frac{\sqrt{1+r^2}}{r}\prt_{\varphi} \Big]\,.
\eea
Each set creates one $SL(2,R)$ algebra as $[L_{+1},L_{-1}]=2L_0\,,\,[L_0,L_{\pm1}]=\mp L_{\pm1}$. 

By inserting $AdS_3$ metric  as a solution, into the NMG equations of motion, we can fix the value of cosmological constant in terms of the mass parameter and the radius of $AdS_3$ 
\be
\Lambda = -\frac{4m^2l^2+1}{4m^2l^4}\,.
\ee
Before we study the linearized equations of motion and find a solution for these equations in a specific background, we must first fix the gauge freedoms. Accordingly, we consider two different gauge fixing conditions and compare their results.

\subsection{The harmonic gauge}
The highest weight solution for the linearized perturbations in harmonic gauge have been discussed extensively in \cite{Li:2008dq},\cite{Liu:2009bk}, in the following global coordinates
\be 
\label{gads} ds^2=l^2\left(-\cosh(\rho)^2 d\tau^2+d\rho^2+\sinh(\rho)^2 d\varphi^2\right)\,.
\ee
But we are interested in another form of the global coordinates as (\ref{adsmetric}), which can be obtained from (\ref{gads}) easily by changing $\sinh(\rho)\rightarrow r$. In addition, we are going to study NMG in different gauge conditions, so as a warm up we review \cite{Li:2008dq},\cite{Liu:2009bk}, in coordinates of (\ref{adsmetric}) and then add more results in the subsequent sections. Meanwhile we compare these results with the asymptotic behavior of propagating solutions. By this technique  we will be able to find new results for other
vacuum solutions of NMG.

$AdS_3$ space-time due to its symmetries makes the linearized equations of motion in (\ref{leomnmg}) simpler. The curvature tensors can be written in terms of the background metric as 
\be \label{syms}
R_{\mu\rho\nu\sigma}=-\frac{1}{l^2}(g_{\mu\nu}g_{\rho\sigma}-g_{\mu\sigma}g_{\rho\nu})\,,\qquad R_{\mu\nu}=-\frac{2}{l^2} g_{\mu\nu}\,,\qquad R=-\frac{6}{l^2}\,.
\ee

If we compute the trace of linearized equations of motion by multiplying equation (\ref{leomnmg}) by  $\bar g^{\mu\nu}$ and impose the harmonic gauge condition $\nabla^\mu h_{\mu\nu}=0$, then we will find a traceless condition for the metric fluctuations, $h={h^{\mu}}_{\mu}=0$. The harmonic gauge and traceless condition simplify the linearized equations of motion (\ref{leomnmg}) into a fourth order differential equation as follow \cite{Li:2008dq}
\be \label{decoup}
(D^{(L)}D^{(R)}D^{(M)}D^{(\tilde{M})}h)_{\mu\nu}=0\,.
\ee
The covariant derivative $D$ is defined as a first order operator 
\be
{{(D^{(L/R)})}_\mu}^\beta=({\delta_\mu}^\beta\pm l{\varepsilon_\mu}^{\alpha\beta}\nabla_{\alpha})\,,\qquad {{(D^{(M/\tilde{M})})}_\mu}^\beta=({\delta_\mu}^\beta\pm \frac{1}{\tilde{\mu}}{\varepsilon_\mu}^{\alpha\beta}\nabla_{\alpha})\,,
\ee
where ${\varepsilon}^{\alpha\beta\gamma}=\frac{1}{\sqrt{g}}{\epsilon}^{\alpha\beta\gamma}$ with ${\epsilon}^{tr\varphi}=1$. For NMG Lagrangian one finds $\tilde\mu$ as \cite{Liu:2009bk},\cite{Myung:2011bn}
\be
\tilde{\mu}=\frac{\sqrt{2+4m^2 l^2}}{2l}\,.
\ee
The differential equation in (\ref{decoup}) describes one left and one right moving massless graviton mode in $AdS_3$. There are also two massive degrees of freedom. These are manifested in the following equations
\bea
&&(D^{(L)}D^{(R)}h)_{\mu\nu}=l^2(\nabla^2-\frac{2}{l^2}) h_{\mu\nu}=0\,,\nn\\
&&(D^{(M)}D^{(\tilde{M})}h)_{\mu\nu}=\frac{1}{\tilde{\mu}^2}(\nabla^2-\frac{2}{l^2}+\mathcal{M}^2) h_{\mu\nu}=0\,,\qquad\mathcal{M}^2=\tilde{\mu}^2-\frac{1}{l^2}=\frac{4m^2l^2-2}{4l^2}\,.
\eea
The mass square relation shows that to maintain the stability (tachyon free condition), the parameters of the theory must  be bounded by values in $m^2 l^2\ge\frac12$  \cite{Myung:2011bn}. 

\subsubsection{The highest weight solutions}
It is possible to solve the linearized equations of motion in (\ref{leomnmg}) by requesting the highest weight conditions. In other words, we can find a solution corresponding to the highest weight representation of the isometry group of $AdS_3$ background. These  solutions are the eigen-modes of $SL(2,R)_L\times SL(2,R)_R$ generators with the following eigen-values
\bea \label{hwc} 
L_{+1} h_{\mu\nu}=0\,,\quad L_0 h_{\mu\nu}=\frac{\omega-k}{2}h_{\mu\nu}\,,\quad\bar{L}_{+1} h_{\mu\nu}=0\,,\quad \bar L_0 h_{\mu\nu}=\frac{\omega+k}{2}\,h_{\mu\nu}\,.
\eea
These conditions lead us to consider the following ansatz for the metric fluctuations
\bea \label{ansatz1}
h_{\mu\nu}=e^{-i(\omega t-k \varphi)}  g(r) \, \left(\begin{array}{ccc} -(1+r^2) f_1(r) & g_1(r) & g_2(r) \\ g_1(r) & (1+r^2)^{-1} f_2(r) & g_3(r) \\ g_2(r) & g_3(r) & r^2\,f_3(r) \end{array} \right)\,.
\eea
By solving the differential equations from the highest weight conditions ${L}_{+1} h_{\mu\nu}=0$ and $\bar{L}_{+1} h_{\mu\nu}=0$, the following behaviors can be obtained for the unknown functions in the ansatz  
\bea\label{f1g3}
&&f_1(r)=\frac{i}{1+{r}^{2}}({C_4}+{C_3}r^2+{C_5}{r}^{4})+\frac12{r}^{2}{C_1}+{C_2}\,,\nn\\
&&f_2(r)=\frac{1}{2{r}^{2}(1+{r}^{2})}\big(({C_1}-4{C_2}-4i{C_3}+2i{C_5}){r}^{4}+(3{C_1}-10{C_2}-6i{C_3}
){r}^{2}+2{C_1}-6{C_2}-4i{C_3}+2i{C_4}\big)\,,\nn\\
&&g_1(r)=\frac{1}{r(1+{r}^{2})}({C_4}+{C_3}{r}^{2}+{C_5}{r}^{4})\,,\qquad g_2(r)=-i{C_5}{r}^{4}-\frac12{r}^{2}(1+{r}^{2}){C_1}+i{C_4}\,,\nn\\
&&g_3(r)=\frac{1}{r(1+{r}^{2})}({C_5}{r}^{4}+(i{C_1}-3i{C_2}+2{C_3}){r}^{2}+i{C_1}-3i{C_2}-{C_4}+2
{C_3})\,,
\eea
where we have used the traceless condition to fix ${f_3}(r)\!=\!-{f_1}(r)\!-\!{f_2}(r)$. In this solution, $C_{i}$'s ($i=1,\dots,5$) are constants of integration and $g(r)$  must be equal to
\be \label{funcg}
g(r)=(1+{r}^{2})^{-\frac{\omega}{2}}{r}^{\pm k}\,,
\ee
where $+$ and $-$ correspond to conditions ${L}_{+1} h_{\mu\nu}=0$ and $\bar{L}_{+1} h_{\mu\nu}=0$ respectively. In  the harmonic gauge the equality of  ${L}_{+1} h_{\mu\nu}=0$ holds when $k=2$ in (\ref{funcg}) and $\bar{L}_{+1} h_{\mu\nu}=0$ holds for $k=-2$ together with $C_1=C_2=C_3=C_5=0$, therefore the metric fluctuations simplify to
\bea
h_{\mu\nu}=\pm C_4 e^{-i(\omega t-k \varphi)}  (1+r^2)^{-\frac{\omega}{2}} \, \left(\begin{array}{ccc}
 i r^2 & -\frac{r}{1+r^2} & \mp ir^2 \\ 
-\frac{r}{1+r^2} & - i \frac{1}{(1+r^2)^2}& \mp\frac{r}{1+r^2} \\ 
\mp ir^2 & \mp\frac{r}{1+r^2} & -i r^2 
\end{array} \right)\,.
\eea
By inserting these fluctuations into the linearized equations of motion, one finds the following values for $\omega$ 
\footnote{These frequencies are consistent with the values obtained in \cite{Liu:2009bk},\cite{Myung:2011bn}.}
\be\label{valomega1}
\omega=0,\,2,\,1\pm\frac12\sqrt{2+4m^2 l^2}\,.
\ee
However there are two constraints here. First,  we expect that the asymptotic fall-off for the metric fluctuations  is  faster than the background metric, and second, we have demanded the tachyon free condition, $m^2l^2\geq\frac{1}{2}$. Therefore if $C_4\neq0$ then $\omega>2$ and the only possible solution is $\omega=1+\frac12\sqrt{2+4m^2 l^2}$.
\subsubsection{The asymptotic behavior of propagating solutions}
To find the asymptotic  behavior of metric perturbations let us  consider the following ansatz
\be \label{ansatz2}
h_{\mu\nu}=l^2 e^{-i(\omega t-k \varphi)} \left( \begin{array}{ccc}
(1+r^2) f_1(r) & g_1(r) & g_2(r) \\
g_1(r) & -(1+r^2)^{-1} f_2(r) & g_3(r)\\
g_2(r) & g_3(r) & -r^2 f_3(r) \end{array} \right)\,. 
\ee
Note that keeping 1 in $1+r^2$ is optional for the asymptotic computations and the final results, as we have checked, will be equal exactly, so for simplicity of calculations we drop it.
In the harmonic gauge, we have three relations among the six unknown functions in $h_{\mu\nu}$. According to the gauge condition these are
\bea\label{gagads}
&&r^4 g'_1(r)+3 r^3 g_1(r)+i  \omega r^2 f_1(r) +i k g_2(r)=0\,,\nn\\
&&-r^2f'_2(r)-2r f_2(r)+i k g_3(r)+i\omega  g_1(r)+r(f_1(r)+f_3(r)) =0\,,\nn\\
&&r^4g'_3(r)+3 r^3 g_3(r)-i k r^2 f_3(r)+i \omega g_2(r) =0\,.
\eea
Consequently we can write $f_1(r), f_3(r)$ and $g_2(r)$ in terms of $g_1(r), g_3(r)$ and $f_2(r)$ and their derivatives. On the other hand  $\bar{g}^{\mu\nu}\delta T^{NMG}_{\mu\nu}=0$. This together with the harmonic gauge condition imply a traceless condition for the metric fluctuations.  We have $f_1(r)\!+\!f_2(r)\!+\!f_3(r)\!=\!0$ accordingly and this constraint fixes another function 
\be\label{tlcads}
g_3(r)=\frac{1}{ik}(r^2 f'_2(r)+3r f_2(r)-i\omega g_1(r))\,.
\ee
If we impose the gauge conditions in (\ref{gagads}) and the constraint in  (\ref{tlcads}) into the linearized equations of motion in (\ref{leomnmg})  then we will find six differential equations for two unknown functions $g_1(r)$ and $f_2(r)$.  Among these equations the $rr$ component is a decoupled differential equation for $f_2(r)$
\bea
&&2r^8f^{(4)}_2+36r^7f'''_2+\big((191-2m^2l^2)r^6-4(k^2-\omega ^2)r^4\big)f''_2+\big((-14m^2l^2+329)r^5+20(\omega ^2-k^2)r^3\big)f'_2\nn\\
&&+2\big((-8m^2l^2+68)r^4+(m^2l^2-\frac{17}{2})(k^2-\omega^2 )r^2+(k^2-\omega^2 )^2\big)f_2=0\,.
\eea
To analyze  this equation and find its asymptotic behavior we will use the Frobenius's method in solving the ordinary differential equations. Let's insert a series solution, $f_2(r)=r^{-B}\sum^\infty_{n=0}\frac{c_n}{r^n}$, into the above equation and take the $r\rightarrow\infty$ limit. Then put the coefficient of the greatest power of $r$ to zero. In this way, the possible values of $B$ are  ($c_0\neq0$)
\be\label{Bf2}
B=2\,, 4\,, 3\pm\frac12\sqrt{2+4m^2l^2}\,.
\ee
By comparing $B$ with frequencies in the highest weight solution obtained in equation (\ref{valomega1}), we see that $B=\omega+2$ for this function.

What about the other unknown function $g_1(r)$?. By a direct computation we observed that one cannot write a linear combination of equations in order to find a decoupled differential equation for $g_1(r)$. To do this, we must bring the higher order derivatives of equations of motion into the game. After doing a little more calculations we obtained an 8th order differential equation for $g_1(r)$. Since it is a very lengthy equation we have not written it and just present the final results. The asymptotic behavior of this 8th order differential equation can be found  by assuming $g_1(r)=r^{-B'}\sum^\infty_{n=0}\frac{c'_n}{r^n}$. The values of $B'$ are given by
\be
B'=1\,, 3\,, 3\,, 5\,, 2\pm\frac12\sqrt{2+4m^2l^2}\,, 4\pm\frac12\sqrt{2+4m^2l^2}\,.
\ee
It is possible to find a relation between $f_2(r)$ and the derivatives of $g_1(r)$. By this relation one can verify that, for each value of $B$ there are two values for $B'$ so that
\be
B'=B\pm1\,.
\ee
We conclude that, if we demand an asymptotic metric fluctuation fall-off similar to the highest weight solutions then $B>4$ and $B'=B-1$. Note that $B'=B+1$ goes more faster to zero. Hence, there are solutions to linearized equations of motion which asymptotically behave similar to the highest weight solutions.
\subsection{$h_{\mu\varphi}=0$ gauge}
In previous section we studied NMG in harmonic gauge. It is interesting to know the behavior of this theory under other gauge fixing conditions. Here we perform all steps in the previous section but for a new gauge $h_{\mu\varphi}=0$. In the appendix C we will show that this gauge can always be attained in $AdS_3$ background, using  an appropriate diffeomorphism.
\subsubsection{The highest weight solutions}
In this gauge, it seems that the equations of motion cannot be written in the decoupled form as (\ref{decoup}). Despite this, we can study different solutions similar to the previous section. 
The metric perturbations ansatz is given in (\ref{ansatz1}). Solving the highest weight equations gives the following functions (note that we have not any traceless condition in this gauge)
\bea
&&f_1(r)=-({C_2}+i{C_4}){r}^{2}-{\frac{i({C_3}-{C_6}+{C_5}+{C_4})}{1+{r}^{2}}}+{C_1}\,,\nn\\
&&f_2(r)=\frac{1}{(1+{r}^{2})}\big(({C_1}+{C_2}-i{C_6}+2i{C_4}-i{C_3}){r}^{4}+(3i{C_4}-2i{C_6}+{C_2}+{C_1}){r}^{2}-i{C_5}\big)\,,\nn\\
&& f_3(r)={r}^{2}{C_2}+i{C_4}{r}^{2}+3i{C_4}+{C_2}+{C_1}+{\frac{i{C_5}}{{r}^{2}}}\,,\quad g_1(r)={\frac{{C_4}{r}^{4}+(2{C_4}+{C_3}){r}^{2}+{C_6}-{C_5}}{r(1+{r}^{2})}}\,,\nn\\ &&g_2(r)=({C_2}+i{C_4}){r}^{4}+(2i{C_4}+{C_2}){r}^{2}-i{C_5}+i{C_6}\,,\quad
g_3(r)={\frac{{C_4}{r}^{4}+{C_5}+{C_6}{r}^{2}}{r(1+{r}^{2})}}\,.
\eea
To impose the gauge condition $h_{\mu\varphi}=0$ in the linearized equations of motion we have to choose $C_1=C_2=C_4=C_5=C_6=0$. Therefore, there are six equations for just one unknown constant $C_3$.
The only consistent solution to these equations is $C_3=0$. Such a solution  is called the pure gauge solution. In another word, in this new gauge to have a highest weight solution with nontrivial values for frequencies and wave numbers, we must choose $f_{i}(r)=g_{i}(r)=0$ for $i=1,2,3$. The same behavior  has been observed previously in \cite{Myung:2011bn} for BTZ black holes.
\subsubsection{The asymptotic behavior of propagating solutions}
Although we showed that in $h_{\mu\varphi}=0$ gauge, the highest weight solution is a pure gauge solution but let us  look at the asymptotic behavior of the metric perturbations. We start from the following ansatz
\be
h_{\mu\nu}=l^2 e^{-i(\omega t-k \varphi)} \left( \begin{array}{ccc}
(1+r^2) f_1(r) & g_1(r) & 0 \\
g_1(r) & -(1+r^2)^{-1} f_2(r) & 0\\
0 & 0 & 0 \end{array} \right)\,. 
\ee
To find the asymptotic behavior, we insert the above ansatz  into the equations of motion (to simplify the calculations we have ignored $1$ in $1+r^2$, the final results are exactly the same). There are five independent equations of motion, which are not enough to decompose the differential equations. On the other hand we have $\bar{g}^{\mu\nu}\delta T^{NMG}_{\mu\nu}=0$ which implies $\delta R=0$. If we use this equation together with its first and second derivatives then we will obtain a decomposed differential equation for $f_1(r)$. By choosing $ml=\xi$ this differential equation is
\begin{align}
&\Big((\frac12\!+\!\xi^2)r^2\!-\!\omega ^2\!+\!k^2\Big)r^{11}f^{(5)}_1\!+\!\Big(9(1\!+\!2\xi^2)r^4\!+\!\big((\frac{41}{2}\!+\!\xi^2)k^2\!-\!20\omega ^2\big)r^2\!-\!\omega ^2k^2\!+\!k^4\Big)r^8f^{(4)}_1\!+\!\Big((47\!+\!93\xi^2\!-\!2\xi^4)r^4&\nn \\
&\!+\!\big(2(59\!+\!\xi^2)k^2\!+\!(4\xi^2\!-\!115)\omega ^2\big)r^2\!-\!4\omega ^2k^2\!+\!6k^4\!-\!2\omega ^4\Big)r^7f'''_1\!-\!2r^4\Big((7\xi^4\!-\!\frac{147}{2}\xi^2\!-\!\frac{77}{2})r^6\!+\!\big((\xi^4\!+\!\frac{23}{2}\xi^2\!-\!104\big)k^2&\nn \\
&\!+\!\omega ^2(103\!-\!13\xi^2)\big)r^4\!+\!2\big((\frac54\!+\!\xi^2)k^2\!-\!3\omega ^2\big)(k^2\!-\!\omega ^2)r^2\!+\!k^2(k^2\!-\!\omega ^2)^2\Big)f''_1\!+\!r^3\Big((\xi^6\!-\!\frac{37}{2}\xi^4\!+\!\frac{179}{4}\xi^2\!+\!\frac{217}{8})r^6\!+\!\big((\xi^4&\nn \\
&\!-\!36\xi^2\!+\!\frac{311}{4})k^2\!-\!3(\xi^2\!-\!\frac{15}{2})(\xi^2\!-\!\frac72)\omega ^2\big)r^4\!+\!\big((\xi^2\!-\!\frac{29}{2})k^2\!-\!3(\xi^2\!-\!\frac72)\omega ^2\big)(k^2\!-\!\omega ^2)r^2\!+\!(k^2\!-\!\omega ^2)^3\Big)f'_1\!\!+\!\!\Big((k^2\!-\!\omega ^2)^3&\nn \\
&\!+\!(\xi^2\!-\!\frac12)^2(\frac12\!+\!\xi^2)r^6\!+\!3(\xi^2\!-\!\frac12)^2(k^2\!-\!\omega ^2)r^4\!+\!3(\frac12\!+\!\xi^2)(k^2\!-\!\omega ^2)^2r^2\Big)k^2f_1=0\,.&
\end{align}
If we consider an asymptotic series solution for the above equation, i.e. $f_1(r)=r^{-B}\sum_{n=0}^\infty \frac{a_n}{r^n}$, then we can read the values of $B$ from equation
$B(B^2-6B+\frac{17}{2}-\xi^2)(B^2-2B+\frac12-\xi^2)=0$. 
So the possible values of $B$ are
\be
B=0\,, B=1\pm\frac12\sqrt{4\xi^2+2}\,, B=3\pm\frac12\sqrt{4\xi^2+2}\,.
\ee
Doing the same computations but for $f_2(r)$, exhibits another complicated differential equation. The asymptotic behavior of $f_2(r)=r^{-B'}\sum_{n=0}^\infty \frac{a'_n}{r^n}$ is reliable when values of $B'$ are
\be
B'=2\,, 1\pm \frac12\sqrt{4\xi^2+2}\,.
\ee 
For $g_1(r)=r^{-B''}\sum_{n=0}^\infty \frac{a''_n}{r^n}$ we find the following values of $B''$
\be
B''=1\,, B''=2\pm \frac12\sqrt{4\xi^2+2}\,, B''=\pm \frac12\sqrt{4\xi^2+2}\,.
\ee
To suppress the blow-up of the metric fluctuations in the asymptotic region, we must restrict the power behaviors to $B>0$, $B'>0$ and $B''>0$.
\section{The warped $AdS_3$ vacuum in NMG}
Another solution to the equations of motion with less symmetries than $AdS_3$ background is $WAdS_3$ vacuum with the following metric in the global coordinates
\be\label{metr}
ds^2=\frac{l^2}{\nu^2+3}\Big[-(1+r^2)d\tau^2+\frac{dr^2}{1+r^2}+\frac{4\nu^2}{\nu^2+3}(d \varphi+rd\tau)^2\Big]\,.
\ee
We define the warp factor as
\be\label{sigma}
\sigma=\frac{2\nu}{\sqrt{\nu^2+3}}\,.
\ee
Since (\ref{metr}) is a stationary metric (its components are independent of the global time $\tau$) then  $\zeta=\partial_\tau$ is a Killing vector of (\ref{metr}) and it can be considered as the generator of time translations.
For $\nu^2<1$ it is always a time-like vector ($|\zeta|=(\frac{l^2}{\nu^2+3}[-1+3(\frac{\nu^2-1}{\nu^2+3})r^2])^{-\frac12}$) but for $\nu^2>1$ there is a transition surface from time-like to space-like. For $0\le\nu^2<1$ the warp factor restricts to $0\le\sigma<1$ and this space-time is called squashed. For $\nu^2>1$ we have $1\le\sigma<2$ and it is called stretched. The special case $\nu=\sigma=1$
corresponds to $AdS_3$ with a fibration \cite{Anninos:2008fx}.

The isometry group of $AdS_3$, i.e. $SL(2,R)_L\times SL(2,R)_R$, is now broken  to $SL(2,R)_L\times U(1)_R$ due to presence of the warp factor. The generators of this symmetry transformation can be constructed out of the Killing vectors as
\bea\label{iso}
L_0=i\partial_\tau\,,\quad 
L_{\pm1}=\pm e^{\pm i\tau}\Big(\frac{r}{\sqrt{1+r^2}}\partial_\tau\mp i\sqrt{1+r^2}\partial_r+\frac{1}{\sqrt{1+r^2}}\partial_\varphi\Big)\,;\quad J=-i\partial_\varphi\,.
\eea
If we  insert (\ref{metr}) into the equations of motion in (\ref{eomnmg}) then we will find the following values for NMG mass parameter $m$ and cosmological constant $\Lambda$ in terms of the warp factor $\sigma$ and  $WAdS_3$ scalar curvature ($R=-\frac{6}{l^2}$)
\be
m^2=-\frac{3}{2l^2}\frac{21\sigma^2-4}{\sigma^2-4}\,,\quad\Lambda=-\frac{3}{2l^2}\frac{21\sigma^4-72\sigma^2+16}{21\sigma^4-88\sigma^2+16}\,.
\ee
Since the warp factor is limited between $0\leq\sigma<2$ we can specify the sign of parameters of the theory for different values of the warp factor.
Suppose that $l^2>0$ (or negative curvature), therefore we find the following domains for the mass  parameter and cosmological constant
\begin{table}[ht]\renewcommand{\arraystretch}{1.5}
\center
\begin{tabular}{|c|c|c|c|c|}\hline
$\sigma$ & $0\le\sigma<\sigma_1$ & $\sigma_1<\sigma<\sigma_2$ & $\sigma_2<\sigma<\sigma_3$ & $\sigma_3<\sigma<2$\\ \hline
$m^2$ & $<0$ & $>0$ & $>0$ & $>0$\\ \hline
$\Lambda$ & $<0$ & $>0$ & $<0$ & $>0$\\ \hline
\end{tabular}
\caption{Behavior of the mass parameter and cosmological constant in NMG for $0\leq\sigma<2$. 
}
\end{table}

The critical values in warp factor which have been appeared in the above table  are as follows
\be
\sigma_1=\frac{2}{\sqrt{21}}\cong0.436\,,\quad\sigma_2=\frac{2\sqrt{189-42\sqrt{15}}}{21}\cong0.489\,,\quad\sigma_3=\frac{2\sqrt{189+42\sqrt{15}}}{21}\cong1.786\,.
\ee
In what follows we would like to find different behaviors of the metric fluctuations around $WAdS_3$ vacuum in different gauge choices.
\subsection{The harmonic gauge}
As we mentioned before, since $WAdS_3$ vacuum has less symmetries than $AdS_3$ we do not have the simple rules like the equations in (\ref{syms}). This makes our equations more complicated, so we just present the final or important results.
\subsubsection{The highest weight solutions}
Unlike $AdS_3$ vacuum we have not a traceless condition here. We consider behavior of the metric fluctuations only in the presence of the harmonic gauge. The highest weight conditions for the metric perturbations are given by the following  relations \cite{Anninos:2009zi}
\be\label{hwe}
J\,h_{\mu\nu}=k\,h_{\mu\nu}\,,\quad{L}_0 h_{\mu\nu}=\omega h_{\mu\nu}\,,\quad{L}_{\pm 1} h_{\mu\nu}=0\,,
\ee
where ${L}_{\pm1}, {L}_0, J$ are Killing vectors that generate the $SL(2,R)_{R}\times U(1)_{L}$ isometry group and are expressed in (\ref{iso}). Like the previous case, we choose our ansatz as \cite{Anninos:2009zi}
\be \label{wans1} 
h_{\mu\nu} (\tau,r,\varphi) = f_3(r) e^{i(k \varphi-\omega \tau)} \, \left(\begin{array}{ccc} f_1(r) & g_1(r) & g_2(r) \\ g_1(r) & f_2(r) & g_3(r) \\ g_2(r) & g_3(r) & C_6 \end{array} \right)\,.
\ee
The $(3,3)$ component of the highest weight condition $L_1 h_{\mu\nu}=0$ fixes the value of $f_3(r)$ to 
\be\label{f6} 
f_3 (r)=e^{k\,\tan^{-1} r}\,(1+r^2)^{-\frac{\omega}{2}}\,.
\ee
By using the above value and by solving the differential equations in other components of $L_1 h_{\mu\nu}=0$ one may hence arrive at
\bea\label{newgauge}
&&g_1(r)=\frac{(2C_1-iC_4)r^2+2(C_2+iC_5-iC_3)r+iC_4}{2(1+r^2)}\,,\,\,
g_2(r)=(C_6-iC_2)r+iC_1\,,\,\,\,
g_3(r)=\frac{C_1 r+C_2}{1+r^2},\nn\\
&&f_1(r)=(C_6-C_3-2iC_2)r^2+(C_4+iC_1)r-C_5\,,\quad f_2(r)=\frac{C_5 r^2+C_4 r+C_3}{(1+r^2)^2}\,.
\eea
In the harmonic gauge we can fix three constants out of all $C_{i}$'s as
\bea\label{wgc123}
{C}_1&=&\frac{i{\sigma}^{2}}{2({k}^{2}+{\sigma}^{4})}\Big({C}_4{k}^{2}+2\big(({\sigma}^{2}-1){C}_3+(1-\omega){C}_5\big)k+{\sigma}^{2}{C}_4(2-\omega)\Big)\,,\nn\\
{C}_2&=&\frac{i{\sigma}^{2}}{2({k}^{2}+{\sigma}^{4})}\Big(2C_3k^2-{C}_4(\sigma^2-(2-\omega))k+2({C}_3-
(1-\omega){C}_5){\sigma}^{2}\Big)\,,\nn\\
{C}_6&=&\frac{1}{2k({k}^{2}+{\sigma}^{4})}\Big(-2{C}_3{k}^{3}+({\sigma}^{2}+2
\omega-3){C}_4{k}^{2}+\Big(\big((-2{C}_5+2
{C}_3)\omega+2{C}_5-4{C}_3\big){\sigma}^{2}\nn\\
&+&2(1-\omega)\big((\omega-1){C}_5+{C}_3\big)\Big)k-{\sigma}^{2}{C}_4(1-\omega)(2-\omega)\Big){\sigma}^{4}\,.
\eea
Now we put (\ref{wans1}) into the linearized equations of motion and simplify these by the gauge conditions. To have a non-trivial solution, the determinant of coefficients of $C_3,C_4$ and $C_5$  for each subset of equations must be zero. Independent of the choice of subsets we always find  two polynomials  $P_1$ and $P_2$.  Defining $\omega=\pm u^\frac12+\frac12$ we have
\bea\label{p1p2}
P_1&=&\frac{1}{64}\big(2(16k^4-136k^2+9)\sigma^8+(64k^6-368k^4+652k^2-45)\sigma^6-4k^2(48k^4-160k^2+95)\sigma^4\nn\\
&+&16k^4(12k^2-19)\sigma^2-64k^6\big)+\frac{1}{16}\big(4(4k^2-5)\sigma^8+(48k^4-152k^2+59)\sigma^6-8k^2(12k^2-17)\sigma^4\nn\\
&+&48k^4\sigma^2\big)u+\big(\frac{1}{2}\sigma^8+\frac{3}{4}(4k^2-5)\sigma^6-3k^2\sigma^4\big)u^2+\sigma^6u^3=0\,,\nn\\
P_2&=&84\sigma^8+24(2k^2-5)\sigma^6-\frac{1}{16}(176k^4+424k^2-333)\sigma^4+\frac{1}{8}(176k^4-208k^2+9)\sigma^2+(11k^4+\frac{9}{2})k^2\nn\\
&+&\big(\sigma^2(48\sigma^4+\frac{31}{2}\sigma^2-5)-2k^2(11\sigma^4-12\sigma^2+1)\big)u-\sigma^2(11\sigma^2-2)u^2=0\,.
\eea
The solution of these equations gives the frequencies of the allowed modes. We will discuss the properties of these modes in the next sections.
\subsubsection{The asymptotic behavior of the propagating solutions}
In order to find behavior of the metric perturbations at the asymptotic limit as propagating modes, let us consider the following ansatz
\be \label{wans2}
h_{\mu\nu}(\tau,r,\varphi)=\frac{l^2(4-\sigma^2)}{12} e^{-i(\omega t-k \varphi)} \left( \begin{array}{ccc}
-(1+r^2) f_1(r) & \sigma^2 g_1(r) & r \sigma^2 g_2(r) \\
\sigma^2 g_1(r) & (1+r^2)^{-1} f_2(r) & \sigma^2 g_3(r)\\
r \sigma^2 g_2(r) & \sigma^2 g_3(r) & -\sigma^2 f_3(r) \end{array} \right)\,. 
\ee
If we impose the harmonic gauge we will find three equations among the unknown functions of the ansatz as
\begin{align}
&(1+r^2)^2f'_2\!-\!i(1+r^2)\Big(\big(k(\sigma^2-1)r^2+\sigma^2\omega r-k\big)g_3-\sigma^2(kr+\omega )g_1-ir(f_1-f_2)\Big)\!-\!r\sigma^2\big((r^2-1)g_2-f_3\big)\!=\!0\,,&\nn\\
&\sigma^2(1+r^2)^2g'_1-i(1+r^2)(kr+\omega)f_1-r\Big(i\big((\sigma^2-1)kr^2+\sigma^2\omega r-k\big)g_2-2\sigma^2(1+r^2)g_1\Big)=0\,,&\nn\\
&\sigma^2(1+r^2)^2g'_3+i\big((kr+\omega )r\sigma^2-k(1+r^2)\big)f_3+\sigma^2r\big(i(kr+\omega )g_2+2(r^2+1)g_3\big)=0\,.&
\end{align}
We may find $f_1(r), f_3(r)$ and $g_2(r)$ from these gauge conditions and insert them into the equations of motion. In this way, we will find six mixed differential equations for three remaining functions. To find the asymptotic $r\rightarrow\infty$ behavior of the solutions there are two approaches. In first approach similar to the previous cases, we can recombine equations of motion and try to find a decoupled differential equation for each unknown function. In second approach, we may consider to use the following behaviors for the remaining unknown functions
\be
g_1(r)\rightarrow C_1r^{-B}\,,\quad f_2(r)\rightarrow C_2r^{-B}\,,\quad g_3(r)\rightarrow C_3r^{-B-1}\,.
\ee
By putting these relations into the equations of motion and by going to the asymptotic region one finds six equations for the three unknown constants $C_1,C_2$ and $C_3$. To find $B$, it is enough to select three out of six equations and then insert the determinant of the coefficients to zero. The values of $B$  in this way must be independent of the choice of equations. Doing all these steps, we will find again the  two polynomials from the highest weight approach in (\ref{p1p2}) just by replacing $\omega$ by $B$.
\subsection{$h_{\mu \varphi}=0$ gauge}
The calculations in this section are roughly analogous to one accomplished for TMG in \cite{Anninos:2009zi} while we do it for NMG. 
\subsubsection{The highest weight solutions}
All steps are similar to the previous gauge. We start from the highest weight conditions in (\ref{hwe}) and then find exactly the same value for $f_3(r)$ as (\ref{f6}).
By substituting this in the remaining equations we will obtain the other functions as follows
\bea \label{wfuncs}
&&{ f_1}(r)=({ C_1}-{ C_6}){r}^{2}+3{ C_2}r-{ C_3}\,,\quad { f_2}(r)={\frac{{ C_3}{r}^{2}+{ C_2}r+{ C_1}}{(1+{r}^{2})^{2}}}\,,\nn\\
&&{ g_1}(r)={\frac{3{ C_2}{r}^{2}+
2({ C_6}-{ C_3})r-{ C_2}}{2i(1+{r}^{2})}},\quad{ g_2}(r)={ C_1}r+{ C_2}\,,\quad{ g_3}(r)={\frac{{ C_2}r-{ C_1}+{ C_6}}{i(1+{r}^{2})}}\,.
\eea
If we use the gauge fixing condition $h_{\mu\varphi}=0$ we will see that $C_6=g_2(r)=g_3(r)=0$ which state that $C_1=C_2=0$. Now we insert these  perturbations into the linearized equations of motion. We can show that for each subset of the equations of motion we have a matrix such that
\be
\mathcal{M}_{3\times3}\left(\begin{array}{c}C_3\\C_4\\C_5\end{array}\right)=0\,.
\ee
The propagating modes are those with $det\mathcal{M} =0$ while the pure gauge modes obtain by $det\mathcal{M}\neq0$ or equivalently $C_3=C_4=C_5=0$. There is a common factor in all determinants of $\mathcal{M}$ for all subsets of equations and we observe that this factor is the $P_2$ polynomial. Therefore, in this gauge only one of the previous polynomials survives. 
This polynomial gives the values of all possible frequencies for each propagating mode in the highest weight as follows
\bea \label{wws}
\omega_{1,2}&\!\!\!\!=\!\!\!\!&\frac12+\frac{1}{\sqrt{11\sigma^2-2}}\Big[96\,\sigma^6+(31-44k^2)\,\sigma^4-(10-48k^2)\sigma^2-4k^2\pm 4\Big(1500\sigma^{12}-1116\sigma^{10}\nn\\&\!\!\!\!+\!\!\!\!&(409+18k^2)\sigma^8-(68+40k^2)\sigma^6+(4+26k^2+k^4)\sigma^4-2(2+k^2)k^2\sigma^2+k^4\Big)^{\frac12}\Big]^{\frac12}\,,\nn\\
\omega_{3,4}&\!\!\!\!=\!\!\!\!&\frac12-\frac{1}{\sqrt{11\sigma^2-2}}\Big[96\,\sigma^6+(31-44k^2)\,\sigma^4-(10-48k^2)\sigma^2-4k^2\pm 4\Big(1500\sigma^{12}-1116\sigma^{10}\nn\\&\!\!\!\!+\!\!\!\!&(409+18k^2)\sigma^8-(68+40k^2)\sigma^6+(4+26k^2+k^4)\sigma^4-2(2+k^2)k^2\sigma^2+k^4\Big)^{\frac12}\Big]^{\frac12}\,.
\eea
\subsubsection{The asymptotic behavior of the propagating solutions}
To find behavior of the metric perturbations in the $h_{\mu \varphi}=0$ gauge let's consider the following ansatz
\be
h_{\mu\nu}= e^{-i(\omega t-k \varphi)} \left( \begin{array}{ccc}
-(1+r^2) f_1(r) & g_1(r) & 0 \\
g_1(r) & (1+r^2)^{-1} f_2(r) & 0\\
0 & 0 & 0 \end{array} \right)\,. 
\ee
Again, in this gauge we have six equations of motion but only five of them are independent. On the other hand we can use the trace of energy-momentum tensor $\bar{g}^{\mu\nu}\delta T_{\mu\nu}=0$. If we use this equation together with its first and second derivatives, we will have eight equations totally. After a little computation we can find three decoupled differential equations for $f_1(r), f_2(r)$ or $g_1(r)$. For each function if we write an asymptotic expansion series then we will find the same behavior. The asymptotic fall-off power is given by the following polynomial, $B=\pm u^\frac12+\frac12$
\bea
P_3&=&84\sigma^8+24(2k^2-5)\sigma^6-\frac{1}{16}(176k^4+424k^2-333)\sigma^4+\frac{1}{8}(176k^4-208k^2+9)\sigma^2+(11k^4+\frac{9}{2})k^2\nn\\
&+&\big(\sigma^2(48\sigma^4+\frac{31}{2}\sigma^2-5)-2k^2(11\sigma^4-12\sigma^2+1)\big)u-\sigma^2(11\sigma^2-2)u^2=0\,.
\eea
As we see $P_3$ is exactly equal to the $P_2$ polynomial in (\ref{p1p2}) in the harmonic gauge.
As another but equivalent approach for finding the asymptotic behavior of the solutions, we can consider the following values for the fluctuation functions
\be
f_1(r)\rightarrow C_1r^{-B}\,,\quad g_1(r)\rightarrow C_2r^{-B}\,,\quad f_2(r)\rightarrow C_3r^{-B}\,.
\ee
By inserting these values into the equations of motion and by going to the large values of $r$ one finds six algebraic equations for the three unknown constants $C_1, C_2$ and $C_3$. The only consistent nontrivial solution is the $P_3$ polynomial. 
\section{Stability}
Two main conditions must be checked in order to have a stable solution. The first one is the positivity of energy in a typical solution and the second one is the reality condition for the frequencies. In this section, we perform both checks to find the domain of reliability of our solutions.
\subsection{Energy condition}
To find the energy of a solution we follow the approach presented in \cite{Anninos:2009zi} for TMG (one may also use the  ADT construction \cite{Abbott:1981ff}).
According to \cite{Maloney:2009ck} associated with each Killing vector $\xi^{\mu}$ of a diffeomorphism invariant theory,  there is a conserved charge $Q(\xi)$  as
\be \label{energy} 
Q(\xi)=\frac{1}{16 \pi G}\int_{\Sigma} \star(\xi^{\mu}E^{(2)}_{\mu\nu}[h^{(1)}] dx^{\nu})\,,
\ee
where $\Sigma$ is a spatial hyper-surface at constant time and $\star$ represents the Hodge star operation. To find the conserved energies,  we substitute the first order perturbations $h^{(1)}_{\mu\nu}$ of the highest weight solutions into the $E^{(2)}_{\mu\nu}$, the energy-momentum pseudo-tensor. 
Consequently, the energy density of a gravitational wave is given by \cite{Anninos:2009zi}
\be \label{egden} {\cal E}=\frac{1}{16\pi G}\int dr \sqrt{-\bar g}\, \bar g^{\tau\mu}E^{(2)}_{\mu\nu}\xi^{\nu}\,,\ee
where we have used $\xi^\mu=(1,0,0)$ to find this density. We consider the physical perturbations in their real form
\be \psi_{\mu\nu}=\alpha\,h_{\mu\nu}+\alpha^{*}\,(h_{\mu\nu})^{*}\,,\ee
where $h_{\mu\nu}$'s are given by (\ref{wans1})-(\ref{newgauge}) in the harmonic gauge and by (\ref{wans1}) together with (\ref{wfuncs}) in the other gauge. After applying the gauge conditions, we can remove the coefficients  $C_1,C_2$ and $C_6$ and write $C_3$ and $C_4$ in terms of $C_5$ from the linearized equations of motion, i.e.,
\be 
C_3=-\frac{S_1(\omega,k,\sigma)}{S_0(\omega,k,\sigma)}\,C_5\,,\quad C_4=\frac{\sigma^2}{k}\,\frac{S_2(\omega,k,\sigma)}{S_0(\omega,k,\sigma)}\,C_5\,,
\ee
where $S_0, S_1$ and $S_2$ are real functions and are given in the appendix A.

To avoid the divergences we consider the energy density per unit length in the $\varphi$\,-\,direction as in \cite{Anninos:2009zi}. The final result for energy can be written as the following sum
\be\label{edens}
{\cal E}=|\alpha C_5|^2\,\sum_{n=0}^{8}\,\left(\int_{-\infty}^{+\infty} dr\,\frac{r^{n}\,e^{2 k\, tan^{-1}r}}{(1+r^2)^{\omega+4}}\right)\Big(B_{n}(k,\sigma)\Big)\equiv\sum_{n=0}^{8}\,A_{n}(k,\sigma)B_{n}(k,\sigma)\,.
\ee
The above integrals are finite for $Re(\omega)>\frac12$ and for $n\leq8$ and obey the following recursion relation \cite{Anninos:2009zi}
\be 
(2\omega+7-n)A_{n}=2k A_{n-1}+(n-1)A_{n-2}\,.
\ee
This relation enables us to write the energy in terms of $A_{0}$, which is real and positive valued. 
As we showed in previous sections, there are two types of solutions labeled by two polynomials: 

\subsubsection*{${\bf{\bullet}}$ Massive modes:}

\begin{figure}[ht]
	\center
		\includegraphics[height=55mm]{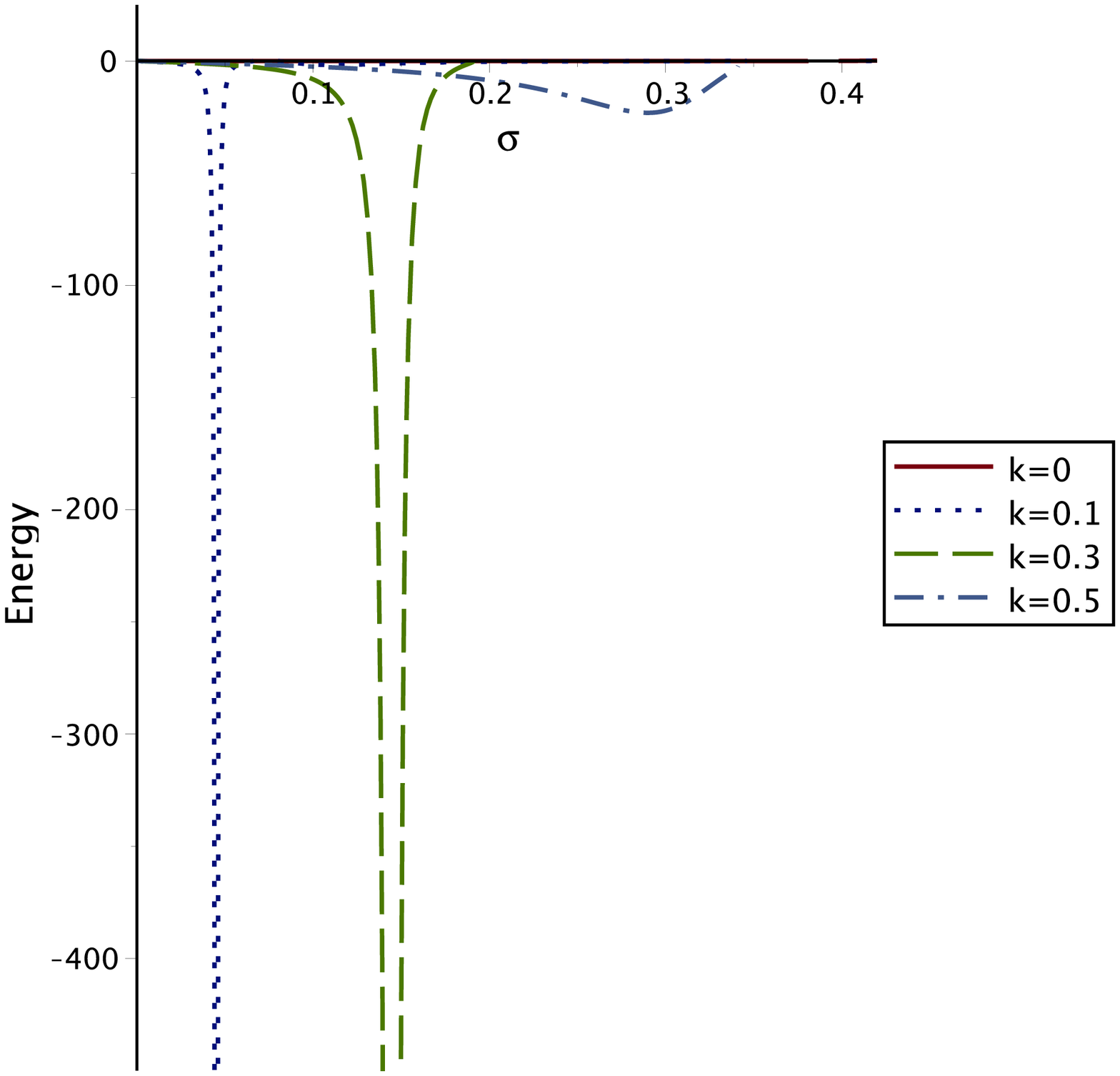}
		\includegraphics[height=55mm]{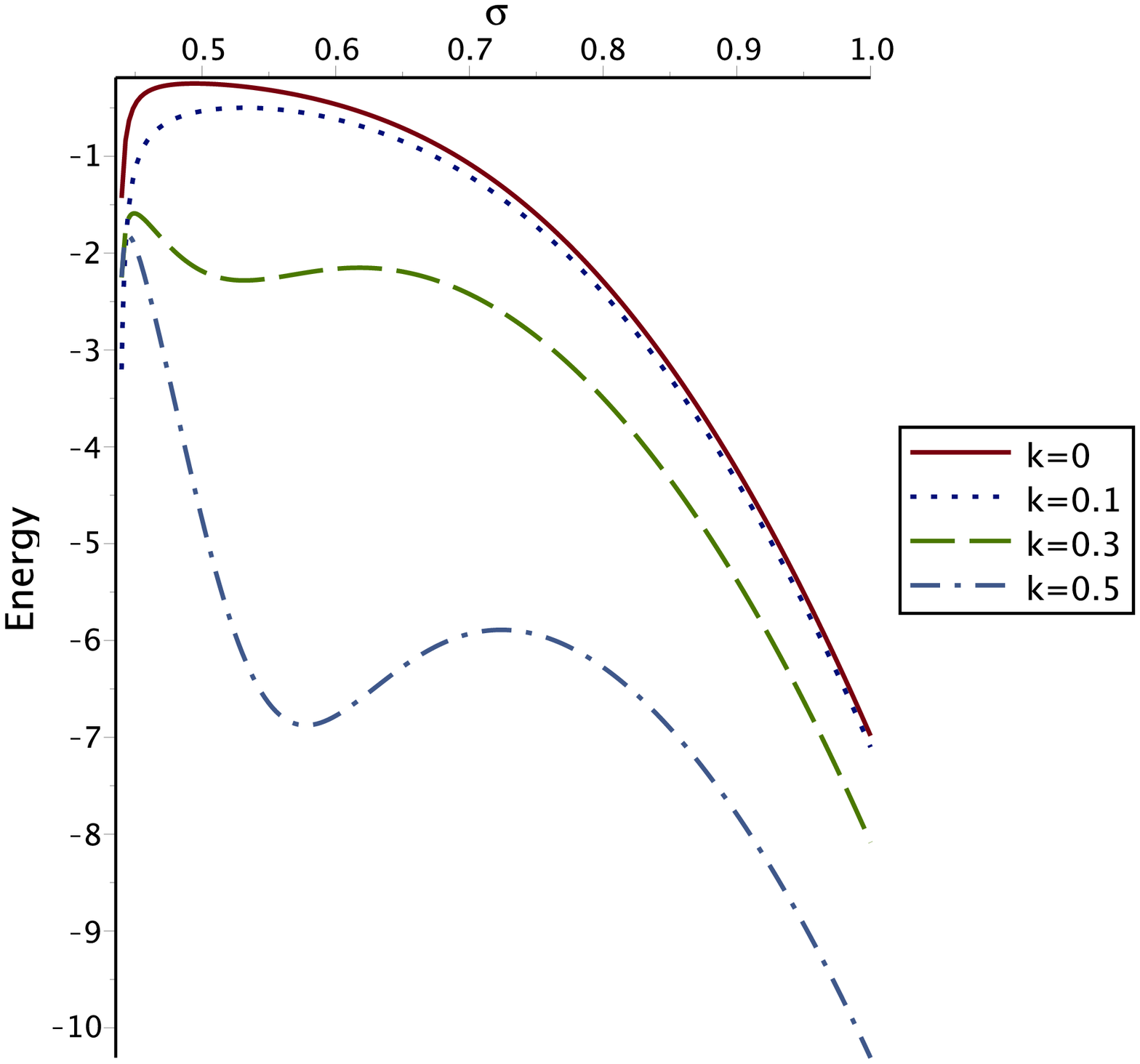}
		\includegraphics[height=55mm]{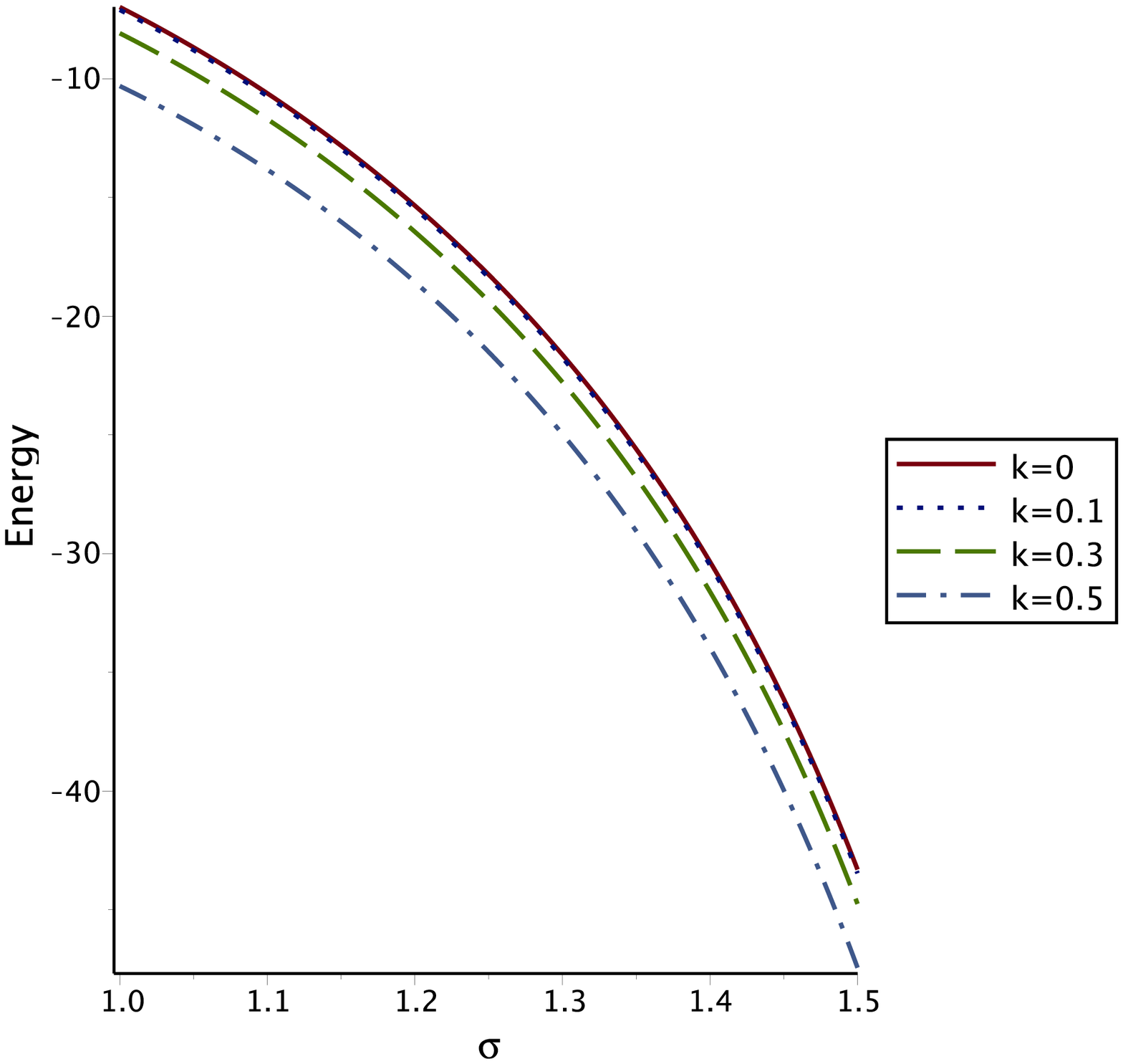}
	\caption{Energy of massive modes in different regions of the allowed parameters for $k=0,0.1,0.3,0.5$.}
	\label{fig:fig1}
\end{figure}

\begin{figure}[ht]
	\center
		\includegraphics[height=65mm]{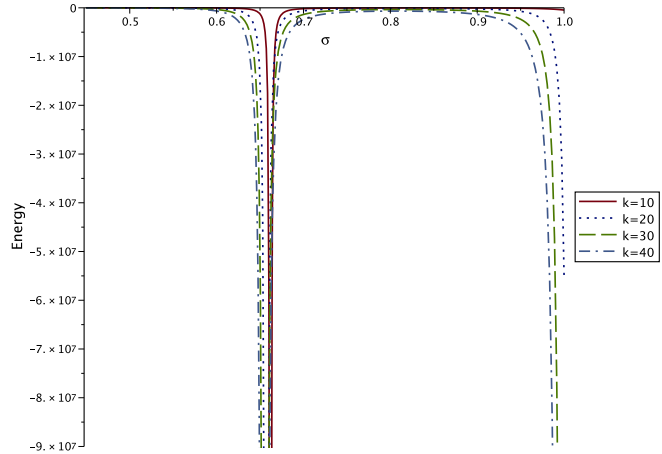}
			\caption{Energy of massive modes in different regions of the allowed parameters for $k=10,20,30,40$.
	}
	\label{fig:fig2}
\end{figure}

If we study the metric fluctuations corresponding to the $P_2$ polynomial, we will observe that the energy density (\ref{edens}) is negative in all the range of $0\leq\sigma<2$ and for all values of $k$, this has been shown in figures \ref{fig:fig1} and \ref{fig:fig2} for different values of $k$. Therefore all massive modes in the squashed or the stretched $WAdS_3$ will make the theory unstable if we cannot exclude these modes in the spectrum.

The asymptotic behavior of the highest weight modes in (\ref{wans1}) is given by
\be\label{asy}
h_{\mu\nu}\sim \left( \begin{array}{ccc}
{\cal O}(r^{2-\omega}) & {\cal O}(r^{-\omega}) & {\cal O}(r^{1-\omega}) \\
&{\cal O}(r^{-2-\omega}) &{\cal O}(r^{-1-\omega}) \\
 &  & {\cal O}(r^{-\omega}) \end{array} \right)\,. 
\ee
This is similar  to TMG  metric fluctuations in \cite{Anninos:2009zi}. In both NMG and TMG the massive propagating modes of $WAdS_3$ do not obey the Comp\'{e}re\,-\,Detournay boundary conditions \cite{Compere:2007in,Compere:2008cv}. 

In general the consistency of boundary conditions requires that the perturbations fall off faster than their corresponding background metric components as we reach the boundary. We suppose that  $WAdS_3$ in NMG has the same boundary conditions as in TMG. This can be expected, since $WAdS_3$ has a very similar behavior both in TMG and NMG. For example the CD boundary conditions has been used in \cite{Ghodsi:2011ua} which leads to the central charges of the dual CFT living at the boundary of $WAdS_3$ space-times. So according to (\ref{metr}) and similar to TMG \cite{Anninos:2009zi} we must only retain those modes which have $\omega(k,\sigma)\leq1$ .

For TMG in \cite{Anninos:2009zi} it has been shown that for stretched $WAdS_3$ all the negative energy propagating modes are excluded from the spectrum and the theory becomes stable. 
In NMG however, we find numerically that in all the interval $0\leq\sigma<2$ and for all values of $k$  there are always modes with $\omega(k,\sigma)>1$. These modes cannot be excluded and make the theory unstable. In figures 3-5, we have sketched $\omega$ for different values of warp factor.

\subsubsection*{${\bf{\bullet}}$ Massless modes:} 
If we examine the energy density of the modes corresponding to the $P_1$ polynomial, the result will be zero. This means that these modes are describing the massless modes of the theory. 
For TMG in \cite{Anninos:2009zi} since they have considered only the $h_{\mu\varphi}=0$ gauge they have not seen these massless modes. To find these modes in TMG we preformed computations of \cite{Anninos:2009zi} in the harmonic gauge. Again, our results contain two types of polynomials. A polynomial for the massive modes which already found in the  $h_{\mu\varphi}=0$ gauge in \cite{Anninos:2009zi} and a new polynomial. The latter has exactly the same polynomial structure as $P_1$ in NMG and its corresponding energy density is zero.

\subsection{Frequency condition}
As we mentioned before, the reality condition of frequencies is another check for the stability of the solutions. To do this we must solve the polynomials and find their roots. These roots were the values of $B$ (the fall-off powers for the propagating solutions) or  the allowed $\omega$ frequencies  for the highest weight modes. We demand that these roots to be positive and real valued. Although the massive modes have negative energies and cannot be excluded but this analysis is needed when one tries to draw the diagrams in figures 1-5. We have presented this analysis in  appendix D.

\section{Extended new massive gravity}
In this section we consider higher curvature corrections to NMG and try to find their effects on
the spectrum of massless and massive perturbative solutions for both $AdS_3$ and $WAdS_3$ background metrics.
The Lagrangian of extended NMG (ENMG) up to third order curvature terms \cite{Gullu:2010pc}-\cite{Sinha:2010ai} is given by
\bea
\label{lagenmg}\mathcal{L}^{ENMG}=\sqrt{-g}\Big(\kappa_3\,{R^{\mu}}_{\,\nu}R^{\nu\rho}R_{\rho\mu}+\kappa_4\,R\,R_{\mu\nu}R^{\mu\nu}+\kappa_5\,R^3\big)\,,
\eea
where
$\kappa_3=-\frac{2}{3m^4},\kappa_4=\frac{3}{4m^4}$ and $\kappa_5=-\frac{17}{96m^4}$.
The equations of motion are obtained by the following energy-momentum tensor
\bea \label{eomc}
T^{ENMG}_{\mu\nu}\!\!\!\!&=&\!\!\!\!\kappa_3\Big(3R_{\mu\alpha}R^{\alpha\beta}R_{\beta\nu}\!-\frac12g_{\mu\nu}{R^{\alpha}}_{\beta}R^{\beta\rho}R_{\rho\alpha}\!+\frac32\big[g_{\mu\nu}\nabla_{\alpha}\!\nabla_{\beta}(R^{\alpha\rho}{R_{\rho}}^{\beta})\!+\Box({R_{\mu}}^{\alpha}R_{\alpha\nu})
-2\nabla_{\alpha}\!\nabla_{(\mu}({R_{\nu)}}^{\beta}{R_{\beta}}^{\alpha})\big]\!\Big)\nn\\
&+&\!\!\!\!\kappa_4\Big(R_{\mu\nu}R_{\alpha\beta}R^{\alpha\beta}+2R{R_{\mu}}^{\alpha}R_{\alpha\nu}\!-\!\frac12g_{\mu\nu}RR_{\alpha\beta}R^{\alpha\beta}+\Box(RR_{\mu\nu})
+\!g_{\mu\nu}\nabla_{\alpha}\nabla_{\beta}(R^{\alpha\beta}R)-2\nabla_{\alpha}\nabla_{(\mu}({R_{\nu)}}^{\alpha}R)\nn\\
&-&\!\!\!\![\nabla_{\mu}\nabla_{\nu}-g_{\mu\nu}\Box](R_{\alpha\beta}R^{\alpha\beta})\Big)
+\kappa_5\Big(3R_{\mu\nu}R^2+3[g_{\mu\nu}\Box-\nabla_{\mu}\nabla_{\nu}]R^2-\frac12g_{\mu\nu}R^3\Big)\,.
\eea
In what follows we will study and solve the linearized form of (\ref{eomc}) around $AdS_3$ and $WAdS_3$ backgrounds. Since the behaviors of ENMG in different gauges are very similar to NMG we present only the results in the harmonic gauge.
\subsection{$AdS_3$ vacuum}
Similar to NMG we can find different properties of $AdS_3$ vacuum in ENMG. In this case, the equations of motion restrict the cosmological constant to
\be
\Lambda=-\frac{8m^4l^4+2m^2l^2+1}{8m^4l^6}\,.
\ee
Once again if we consider perturbations around $AdS_3$ vacuum we will find a  fourth order differential equation similar to (\ref{decoup}) but in this case
\be
\tilde{\mu}=\frac{1}{2l}\sqrt{\frac{8m^4l^4+4m^2l^2+3}{2m^2l^2+1}}\,.
\ee
The differential equation (\ref{decoup}) describes again a massless graviton mode and a massive graviton with mass square
\be
\mathcal{M}^2=\frac{8m^4l^4-4m^2l^2-1}{4l^2(2m^2l^2+1)}\,,
\ee
so the tachyon free condition occurs for $m^2l^2\ge\frac{1+\sqrt{3}}{4}$ or $-\frac12\le m^2l^2\le\frac{1-\sqrt{3}}{4}$.

The highest weight solutions can be found by inserting the ansatz (\ref{ansatz1}) into the highest weight equations (\ref{hwc}) and determining $h_{\mu\nu}$. Substituting these values into the linearized equations of motion accompanied with the gauge conditions give the following values for frequencies
\be
\omega=0,\,2,\,1\pm \frac{1}{2}\sqrt{\frac{8m^4l^4+4m^2l^2+3}{2m^2l^2+1}}\,.
\ee
Comparing these results with (\ref{valomega1}) shows that only the massive mode gets correction.

We can confirm the above result by looking to the asymptotic behavior of the propagating solutions by inserting the ansatz (\ref{ansatz2}) into the linearized equations of motion and using the gauge conditions. As an example we obtain the following differential equation for $f_2(r)$ 
\bea
0&\!\!\!\!=\!\!\!\!&(m^{2}l^{2}+\frac12)r^{8}f^{(4)}_2+18(m^{2}l^{2}+\frac12)r^{7}f^{(3)}_2
-\big((m^{4}l^{4}-{\frac{191}{2}}m^{2}l^{2}-{\frac{381}{8}})r^{2}+2(m^{2}l^{2}+\frac12)(k^2-\omega^2)\big)r^{4}f''_2\nn\\
&\!\!\!\!-\!\!\!\!&\big((7m^{4}l^{4}-{\frac{329}{2}}m^{2}l^{2}-{\frac{651}{8}})r^{2}+10(m^{2}l^{2}+\frac12)(k^2-\omega^2)\big)r^{3}f'_2+\big((-8m^{4}l^{4}+68m^{2}l^{2}+33)r^{4}\nn\\
&\!\!\!\!+\!\!\!\!&(m^{4}l^{4}-\frac{17}{2}m^{2}l^{2}-{\frac{33}{8}})(k^2-\omega^2)r^{2}+(m^{2}l^{2}+\frac12)(k^2-\omega^2)^{2}\big)f_2\,.
\eea
The leading term in the series solution around the boundary behaves as $r^{-B}$ such that $B$ has the following values
\be
B=2\,,4\,,3\pm\frac12\sqrt{\frac{8m^4l^4+4m^2l^2+3}{2m^2l^2+1}}\,.
\ee
The relation between $\omega$ in the highest weight mode and $B$ in the propagating mode is exactly similar to  NMG i.e. $B=\omega+2$. 
\subsection{Warped-$AdS_3$ vacuum}
If we consider the NMG Lagrangian and its curvature corrections and insert the vacuum solution (\ref{metr}) into the equations of motion $T^{NMG}_{\mu\nu}+T^{ENMG}_{\mu\nu}=0$, then we will find the following values for the mass parameter and the cosmological constant ($\Delta=\sqrt{171\sigma^4+264\sigma^2-80}$)
\bea
&&m^2=m^2_{\pm}=-\frac{3}{4l^2}\frac{21\sigma^2-4\pm\Delta}{\sigma^2-4}\,,\nn\\
&&\,\,\,\,\Lambda=\Lambda_{\pm}=\frac{-783\sigma^6+2340\sigma^4
+1008\sigma^2-576\pm(-63\sigma^4+216\sigma^2-48)\Delta}{(\sigma^2-4)(21\sigma^2-4\pm\Delta)^2l^2}\,.
\eea
By considering $l^2>0$ the following behaviors for the mass parameter and the cosmological constant will be obtained (see tables 2 and 3). The reality of cosmological constant constraints the warp factor to begin from $\sigma_c=\frac{2\sqrt{-627+342\sqrt{6}}}{57}\cong0.509$\,.
\begin{table}[ht]\renewcommand{\arraystretch}{1.5}
\center
\begin{tabular}{|c|c|c|c|}\hline
$\sigma$&$\sigma_c\le\sigma<\sigma_L$&$\sigma_L<\sigma<\sigma_R$&$\sigma_R<\sigma<2$\\\hline
$m_{+}^2$&$>0$&$>0$&$>0$\\\hline
$\Lambda_{+}$&$>0$&$<0$&$>0$\\\hline
\end{tabular}
\caption{Behaviors of $m_{+}$ and $\Lambda_{+}$ for $\sigma_c\leq\sigma<2$\,.}
\end{table}

In table 2, $\sigma_L$ and $\sigma_R$ are two real roots (between 0 and 2) of $\Lambda_+=0$. Numerically they are equal to $\sigma_L\cong0.558$ and $\sigma_R\cong1.802$.

\begin{table}[ht]\renewcommand{\arraystretch}{1.5}
\center
\begin{tabular}{|c|c|c|c|}\hline
$\sigma$&$\sigma_c\le\sigma<\frac{2}{\sqrt{15}}$&$\frac{2}{\sqrt{15}}<\sigma<\frac{2}{\sqrt{3}}$&$\frac{2}{\sqrt{3}}<\sigma<2$\\\hline
$m_-^2$&$>0$&$<0$&$>0$\\\hline
$\Lambda_-$&$>0$&$>0$&$<0$\\\hline
\end{tabular}
\caption{Behaviors of $m_{-}$ and $\Lambda_{-}$ for $\sigma_c\leq\sigma<2$\,.}
\end{table}
\subsubsection{Massless and massive modes}
Similar to NMG we use the ansatz in (\ref{wans1}) and put it into the equations in (\ref{hwe}). We achieve the same results as (\ref{newgauge}) and (\ref{wgc123}). If we substitute these results into the linearized equations of motion then we will find non-trivial solutions when the determinant of coefficients is zero. Similar to NMG  in the harmonic gauge, here we have two polynomials $P'_1$ and $P'_2 $. The first polynomial is $P'_1=P_1$. In other words, we find again the massless mode of (\ref{p1p2}) and it does not receive any correction. The other polynomial, which represents the massive modes can be written as
\be \label{p2enmg} 
P'_2=E_0(k,\sigma)+E_1(k,\sigma)\,u+E_2(k,\sigma)\,u^2+E_3(k,\sigma)\,u^3=0\,,
\ee
where $\omega=\pm u^{\frac12}+\frac12$ and the functions $E_{i}$ are given in appendix B.

If we try to find the asymptotic behavior of propagating solutions, we will obtain the above results exactly.
\section{Summary and Conclusions}
In this paper we have mainly discussed about the stability of $AdS$ and warped $AdS$ vacua in new massive gravity. First we found the equations of motion for NMG and then linearized around an arbitrary background $\bar g_{\mu\nu}$. We determined the value of cosmological constant for each solution in terms of other parameters in the theory such as $m,l$ and $\sigma$. 


In this paper we considered the behavior of metric perturbations from two points of view. In the first view, the metric fluctuations are solutions for the highest weight conditions as well as the linearized equations of motion. In the second view, the equations of motion decomposed into some differential equations for each component. That could be solved by analytical methods in the asymptotic limit. We have used two different gauge conditions to write the linearized equations of motion. 

In $WAdS_3$ vacuum the value of frequency for a highest weight mode was related to the radial fall-off power parameter of the general propagating modes at the boundary. In fact, in most cases that we have studied, the propagating modes of the metric fluctuations were also belonged to the representations of isometry group of the background. 

We observed that the existence of a mode in a vacuum, depends on the gauge choice. For example in $WAdS_3$ vacuum and in the harmonic gauge we obtain two polynomials, $P_1$ and $P_2$, while in $h_{\mu \varphi}=0$ gauge only $P_2$ exists. This can be seen for $AdS_3$ vacuum as well. In the harmonic gauge there is a highest weight mode while it becomes pure gauge when we go to $h_{\mu\varphi}=0$ gauge. These behaviors back to the fact that the harmonic gauge does not completely fix the gauge redundancy therefore we see the massless modes in this gauge.
But $h_{\mu \varphi}=0$ fixes the gauge completely and the massless modes become invisible in this gauge.

In this paper we discuss about the stability of the vacuum perturbations at the asymptotic limit and find the domains of validity for parameters in different gauges. 
We show that $P_1$ polynomial describes the massless modes of the theory. For the squashed warped space-time there is always one possible allowed frequency but for the stretched warped space-time the stability is limited to some regions of space of parameters. We also show that the massive modes which describe by $P_2$ polynomial have always negative energies both in the squashed and in the stretched warped space-time. 

By looking at the asymptotic behavior of the massive propagating modes and by analogy with TMG \cite{Anninos:2009zi} we  can try to exclude the negative energy modes based on the CD boundary conditions. We observe  that unlike the TMG we cannot exclude these modes from the spectrum and therefore these modes make the theory unstable. 

We have also considered the extension of NMG, which are constructed from curvature terms and are consistent with the $AdS/CFT$ context. This extended Lagrangian did not change the main results but only corrected the values of mass for the massive propagating modes of $AdS_3$ background. 
For warped space-time we again showed that there are two polynomials. $P'_2$ which describes the massive modes and $P'_1$ which is exactly equal to $P_1$ and therefore the massless modes do not correct by the higher curvature terms. 

We also looked at the TMG model in the harmonic gauge. We found two polynomials, one was exactly $P_1$ in NMG and the other one was the result of \cite{Anninos:2009zi} for TMG in $h_{\mu\varphi}=0$ gauge.

\section*{Acknowledgment}
A. G. would like to thanks D. Anninos, M. Guica and A. E. Mosaffa for very useful discussions. D. M. would like to thanks H. Golchin for discussions.
This work was supported by Ferdowsi University of Mashhad under the grant 2/23391 (02/08/1391).
\appendix
\section{$S_0, S_1, S_2$ functions}
{\bf{$\bullet$ The real functions in $h_{\mu\varphi}=0$ gauge}}
\bea \begin{aligned}
S_0&=42(\omega\!-\!1){\sigma}^{10}\!+\!( 15 {\omega}^{3}\!-\!45 {\omega}^{2}\!+\! 3( 5 {k}^{2}\!+\!1 ) \omega\!+\!48\!+\!6 {k}^{2} ) {\sigma}^{8}\!-\!(2 {\omega}^{5}\!+\!7 {\omega}^{4}\!+\! ( 4 {k}^{2}\!-\!54 ) {\omega}^{3}\!+\!(9 {k}^{2}\!+\!68 ) {\omega}^{2}\nn\\
&\!+\!( 2 {k}^{4}\!-\!44 {k}^{2}\!-\!8
 ) \omega\!+\!50 {k}^{2}\!+\!2 {k}^{4} ) {\sigma}^{6}\!+\! ( 2 {\omega}^{6}\!-\!
10 {\omega}^{5}\!+\! ( 4 {k}^{2}\!+\!16 ) {\omega}^{4}\!-\!(8 {k}^{2}\!+\!8) {\omega}^{3}\!+\! ( 2 {k}^{4}\!+\!17 {k}^{2} ) {\omega}^{2}\nn\\
&\!+\!(2 {k}^{4}\!-\!57 {k}^{2}) \omega\!+\!8 {k}^{4}\!+\!56 {k}^{2} ) {\sigma}^{4}
\!-\!2 {k}^{2} ( 4\!+\!{k}^{2}\!-\!4 \omega\!+\!{\omega}^{2} )  ( {k}^{2}\!-\!\omega\!+\!2 {\omega}^{2}) {\sigma}^{2}\!+\!2 {k}^{4} ( 4\!+\!{k}^{2}\!-\!4 \omega\!+\!{\omega}^{2} ) ,\\
S_1&=42(\omega\!-\!1) {\sigma}^{10}\!+\! ( 15 {\omega}^{3}\!+\!27 \omega\!+\!15 {k}^{2}\omega\!-\!51 {\omega}^{2}\!+\!24 ) {\sigma}^{8}\!-\!( 2 {\omega}^{5}\!-\!10 {\omega}^{4}\!+\!( 4 {k}^{2}\!+\!31 ){\omega}^{3}\!-\!( 25 {k}^{2}\!+\!68 ) {\omega}^{2}\nn\\
&\!+\!( 2 {k}^{4}\!+\!41 {k}^{2}\!+\!60) \omega\!-\!18 {k}^{2}\!-\!15 {
k}^{4} ) {\sigma}^{6}\!+\! ( 2 {\omega}^{5}\!-\! (2 {k}^{2}\!+\!10) {\omega}^{4}\!+\!( 16 {k}^{2}\!+\!16  {\omega}^{3}\!-\! (8\!+\!4 {k}^{4}\!+\!40 {k}^{2} ) {\omega}^{2}\nn\\
&\!+\!( 51 {k}^{2}\!+\!14 {k}^{4}
 ) \omega\!-\!36 {k}^{2}\!-\!23 {k}^{4}\!-\!2 {k}^{6} ) {\sigma}^{4}
\!+\!2 {k}^{2}( 4\!+\!{k}^{2}\!-\!4 \omega\!+\!{\omega}^{2} ) ((2{k}^{2}\!+\! {\omega}
^{2}\!-\!2\omega) {\sigma}^{2}\!-\!{k}^{2}) ,\\
S_2&=42({k}^{2}\!-\! {\omega}(\omega\!-\!2)) {\sigma}^{8}\!-\!(15 {\omega
}^{4}\!-\!45 {\omega}^{3}\!-\!6 {\omega}^{2}\!+\! (57 {k}^{2}\!+\!72) \omega\!-\!36 {k}
^{2}\!-\!15 {k}^{4} ) {\sigma}^{6}\!+\!( 2 {\omega}^{6}\!-\!8 {\omega}^{5}\!+\!( 6\!+\!2 {k}^{2} ) {\omega}^{4}\nn\\
&\!+\! (8\!-\!26 {k}^{2} ) {\omega}^{3}\!+\! ( 45 {k}^{2}\!-\!2 {k}^{4}\!-\!8 ) {\omega}^{2}\!+\!( 40 {k}^{2}\!-\!18 {k}^{4} ) \!-\!2 {k}^{6}\!+\!19 {k}^{4}\!-\!60 {k}^{2} ) {\sigma}^{4}\omega\!+\!4  ( {\omega}^{5}\!-\!5 {\omega}^{4}\!+\! ( 7\!+\!2 {k}^{2} ) {\omega}^{3}\nn\\
&\!+\! ( \frac32\!-\!4 {k}^{2} ) {\omega}^{2}
\!+\!(\frac72 {k}^{2}\!-\!7\!+\!{k}^{4} ) \omega\!+\!{k}^{4}\!-\!2 {k}^{2} ) {k}^{2}{\sigma}^{2}\!-\!4
  ( \omega\!+\!\frac12 )( 4\!+\!{k}^{2}\!-\!4 \omega\!+\!{\omega}^{2} ) {k}^{4}\,,
\end{aligned}\eea
{\bf{$\bullet$ The real functions in the harmonic gauge}}
\bea\begin{aligned}
S_0&=(42 {k}^{4} \!-\!(42 {\omega}^{2} \!-\!420 \omega\!+\!336) {k}^{2} \!-\!84
 {\omega}^{4} \!+\!84 \omega \!-\!294 {\omega}^{2} \!+\!294 {\omega}^{3} ) {\sigma}^{14} \!+\! ( 78
 {k}^{6}\!+\! ( 99 {\omega}^{2}\!+\!417 \omega\!-\!420 ) {k}^{4} \nn\\
&\!-\!(36 {\omega}^{4} \!-\!624 {\omega}^{3} \!+\!630 {\omega}^{2} \!+\!996 \omega \!-\!1056 ) {k}^{2} \!-\!57 {\omega}^{6} \!+\!606 {\omega}^{2} \!-\!117 {\omega}^{4} \!-\!375 {\omega}^{3} \!+\!207 {\omega}^{5}\!-\!264 \omega){\sigma}^{12}\!-\!(10 {k}^{8}\nn\\
&\!+\! (28 {\omega}^{2}\!-\!200 \omega\!+\!296){k}^{6}\!-\!(24 {\omega}^{4}\!-\!398 {\omega}^{3}\!+\!542 {\omega}^{2}\!+\!1142 \omega\!-\!1034 ) {k}^{4}\!-\!(4 {\omega}^{6} \!-\!196 {\omega}^{5} \!+\!266 {\omega}^{4} \!+\!900 {\omega}^{3}\nn\\
& \!-\!1368 {\omega}^{2} \!-\!548 \omega \!+\!1104) {k}^{2} \!+\!2 {\omega}^{8} \!-\!2 {\omega}^{7} \!-\!20 {\omega}^{6} \!-\!2 {\omega}^{5} \!+\! 160 {\omega}^{4}\!-\!134 {\omega}^{3} \!-\! 196 {\omega}^{2} \!+\! 192 \omega ) {\sigma}^{10} \!+\!( 4 {k}^{10} \!+\!  ( 16 {\omega}^{2}\nn\\
& \!-\!32 \omega \!+\!30) {k}^{8} \!+\! (24 {\omega}^{4} \!-\! 100 {\omega}^{3} \!+\!168 {\omega}^{2} \!-\! 462 \omega \!+\! 398) {k}^{6}\!+\!(16 {\omega}^{6} \!-\! 108 {\omega}^{5} \!+\! 270 {\omega}^{4} \!-\!664 {\omega}^{3}\!+\!497 {\omega}^{2} \!+\!1021 \omega  \nn\\
&\!-\!912 ) {k}^{4} \!+\! (4 {\omega}^{8} \!-\!44 {\omega
}^{7} \!+\!156 {\omega}^{6} \!-\!242 {\omega}^{5} \!+\!56 {\omega}^{4}\!+\!450 {\omega}^{3} \!-\!606 {\omega}^{2} \!-\!56 \omega \!+\!384 ) {k}^{2} \!-\!4 {\omega}^{2} ( \omega^2 \!-\!1 ) ^{2}( \omega \!-\!2 ) ^{3} ) {\sigma}^{8} \nn\\
&\!-\! (16 {k}^{8} \!+\! ( 52 {\omega}^{2} \!-\!112 \omega \!+\!46
 ) {k}^{6} \!+\! (60 {\omega}^{4}\!-\!264 {\omega}^{3}\!+\!348 {\omega}^{2}\!-\!324 \omega\!+\!
156){k}^{4}\!+\!(28 {\omega}^{6}\!-\!192 {\omega}^{5}\!+\!426 {\omega}^{4}\!-\!318 {\omega}^{3}\nn\\
&\!-\!154 {\omega}^{2}\!+\!408 \omega\!-\!192 ) {k}^{2}
\!+\!4 \omega ( \omega^2\!-\!1 )( {\omega}^{3}\!-\!6 {\omega}^{2}\!+\!4 \omega\!+\!6 )(\omega\!-\!2
 ) ^{2} ) {k}^{2}{\sigma}^{6}\!+\!(24{k}^{6}\!+\! (60 {\omega}^{2}\!-\!144 \omega\!+\!58){k}^{4}\nn\\
&\!+\!(48 {\omega}
^{4}\!-\!228 {\omega}^{3}\!+\!304 {\omega}^{2}\!-\!62 \omega\!-\!88){k}^{2}\!+\!4  (3{\omega}^{4}\!-\!9 {\omega}^{3}\!-\!3{\omega}^{2}\!+\!11 \omega\!+\!4
 )  ( \omega\!-\!2 ) ^{2} ) {k}^{4}{\sigma}^{4}\!+\!4  ( {k}^{2}\!+\! ( \omega\!-\!2 ) ^{2}
 ) {k}^{8}\nn\\
&\!-\!4  ( {
k}^{2}\!+\! ( \omega\!-\!2 ) ^{2} )(4{k}^{2}\!+\!3 {\omega}^{2}\!-\!4(\omega
\!+\!1) ) {k}^{6}{\sigma}^{2}\,,\\
S_1&=(84 {k}^{4} \!-\!168 {k}^{2}  \!-\!42 {\omega}^{4} \!+\!84 {\omega}^{3} \!+\! ( 42 {k}^{2} \!+\!42 ) {\omega}^{2}\!+\! ( 210 {k}^{2} \!-\!84 ) \omega) {\sigma}^{14} \!-\!(78 {\omega}^{6} \!-\!333 {\omega}^{5} \!+\!(99 {k}^{2} \!+\!522){\omega}^{4}\nn\\
&\!-\!(876 {k}^{2}\!+\!537 ) {\omega}^{3}\!-\!(36 {k}^{4} \!-\!1443 {k}^{2}\!-\!534) {\omega}^{2}\!-\!(543 {k}^{4}\!+\!600 {k}^{2}\!+\!264 ) \omega\!-\!57 {k}^{6}\!+\!828 {k}^{4}\!+\!240 {k}^{2} ) 
{\sigma}^{12}\nn\\
&\!+\!(10 {\omega}^{8}\!-\!58 {\omega}^{7}\!+\! ( 28 {k}^{2}\!+\!80 ) {\omega}^{6}\!+\! ( 28 {k}^{2}\!+\!110 ) {\omega}^{5}\!+\! (24 {k}^{4}\!-\!196 {k}^{2}\!-\!292){\omega}^{4}\!+\! (230
 {k}^{4}\!-\!22 {k}^{2}\!-\!22) {\omega}^{3}\nn\\
&\!+\!(4 {k}^{6} \!-\!702 {k}^{4} \!+\!270 {k}^{2} \!+\!364) {\omega}^{2} \!+\! (144 {k}^
{6} \!-\!376 {k}^{4} \!-\!588 {k}^{2} \!-\!192) \omega \!-\!2 {k}^{8} \!-\!426 {k}^{6} \!+\!860 {k}^{4} \!+\!792 {k}^{2}){\sigma}^{10}\nn\\
&\!-\!(4 {\omega}^{10}\!-\!28 {\omega}^{9}\!+\!(16 {k}^{2}\!+\!64) {\omega}^{8}\!-\! ( 84 {k}^{2}\!+\!24 ) {\omega}^{7}\!+\!(24 {k}^{4}\!+\!154 {k}^{2}\!-\!108) {\omega}^{6}\!-\!(84 {k}^{4}\!+\!114 {k}^{2}\!-\!132) {\omega}^{5}\nn\\
&\!+\!(16 {k}^{6} \!+\!132 {k}^{4} \!-\!14 {k}^{2} \!+\!8) {\omega}^{4} \!-\!( 28 {k}^{6} \!-\!144 {k}^{4} \!-\!294 {k}^{2} \!+\!80){\omega}^{3} \!+\!(4 {k}^{8} \!+\!58 {k}^{6} \!-\!518 {k}^{4} \!-\!474 {k}
^{2} \!+\!32 ) {\omega}^{2}\nn\\
&\!+\!(234 {k}^{6}\!+\!191 {k}^{4}\!-\!72 {k}^{2}) \omega\!+\!16 {k}^{8}\!-\!585 {k}^{6}\!-\!124 {k}^{4}\!+\!384 {k}^{2}){\sigma}^{8}\!+\!(16 {\omega}^{8}\!-\!100 {\omega}^{7}\!+\!4( 13 {k}^{2}\!+\!48
 ) {\omega}^{6}\nn\\
 &\!-\!4(57 {k}^{2}\!+\!5) {\omega}^{5}\!+\! (60 {k}^{4}\!+\!290 {k}^{2}\!-\!304 ) {\omega}^{4}\!-\!(156 {k}^{4}\!-\!6 {k}^{2}\!+\!216) {\omega}^{3}\!+\!(28 {k}^{6}\!+\!136 {k}^{4}\!-\!194 {k}^{2}\!+\!96) {\omega}^{2}\nn\\
&\!-\!4(7 {k}^{6} \!-\!32 {k}^{4} \!-\!38 {k}^{2} \!+\!24)\omega \!+\!(4{k}^{6} \!+\!38 {k}^{4} \!-\!232 {k}^{2} \!-\!176){k}^{2}){k}^{2}{\sigma}^{6} \!-\!(24 {\omega}^{6} \!-\!132 {\omega}
^{5} \!+\! 4( 15 {k}^{2} \!+\!52 ) {\omega}^{4}\nn\\
&\!-\!4(51 {k}^{2}\!-\!5) {\omega}^{3}\!+\! (48 {k}^{4}\!+\!178 {k}^{2}\!-\!256) {\omega}^{2}\!-\!(72 {k}^{4}\!-\!54 {k}^{2}\!+\!80) \omega\!+\!12{k}^{6}\!+\!36 {k}^{4}\!-\!80 {k}^{2}\!+\!64) {k}^{4}{\sigma}^{4}\nn\\
&\!+\!4(3{k}^{2}\!+\!4 {\omega}^{2}\!-\!3\omega\!-\!4)({k}^{2}\!+\!4\!-\!4 \omega\!+\!{\omega}^{2} ) {k}^{6}{\sigma}^{2}\!-\!4 {k}^{8} ( {k}^{2}\!+\!4\!-\!4 \omega\!+\!{\omega}^{2})\,,\\
S_2&=( 252 {\omega}^{3} \!-\! 252 {\omega}^{2} \!+\! ( 252 {k}^{2} \!-\! 252 ) \omega \!-\! 504 {k}^{2} ) {\sigma}^{12} \!+\!  ( 270 {\omega}^{5} \!-\! 330 {\omega}^{4} \!+\!(540 {k}^{2} \!-\! 702 ) {\omega}^{3} \!+\!  ( 426 \!-\! 1080 {k}^{2}){\omega}^{2}\nn\\
&\!+\!(744\!+\!270 {k}^{4} \!-\!516 {k}^{2} ) \omega \!+\!984 {k}^{2} \!-\!750 {k}^{4} ) {\sigma}^{10} \!-\!(24 {\omega}^{7} \!-\!262 {\omega}^{6} \!+\!(712 \!+\!72 {k}^{2}){\omega}^{5} \!+\! ( 270 \!+\!382 {k}^{2} ) {\omega}^{4}\!-\!(72 {k}^{4}\nn\\
&\!+\!1564 {k}^{2}\!-\!868){\omega}^{3}\!-\!(22 {k}^{4}\!-\!1434 {k}^{2}\!+\!340) {\omega}^{2}\!-\!(24 {k}^{6}\!+\!852 {k}^{4}\!-\!992 {k}^{2}\!+\!480 ) \omega\!+\!1652 {k}^{4}\!-\!456 {k}^{2}\!-\!142 {k}^{6} ) {\sigma}^{8}\nn\\
&\!+\! ( 8 {\omega}^{9}\!-\!52 {\omega}^{8}\!+\!
 ( 108\!+\!32 {k}^{2} ) {\omega}^{7}\!-\!(152 {k}^{2}\!+\!40){\omega}^{6}\!+\!(48 {k}^{4}\!+\!216 {k}^{2}\!-\!116) {\omega}^{5}\!-\!(144 {k}^{4}\!+\!340 {k}^{2}\!-\!96) {\omega}^{4}\!+\!(32 {k}^{6}\nn\\
&\!-\!108 {k}^{4}\!+\!892 {k}^{2}\!+\!36) {\omega}^{3}\!-\!(40 {k}^{6}\!-\!40 {k}^{4}\!+\!356 {k}^{2}\!+\!40) {\omega}^{2}\!+\!(8 {k}^{8}\!+\!954 {k}^{4}\!-\!668 {k}^{2}) \omega\!+\!4 {k}^{8}\!+\!340 {k}^{6}\!-\!1086 {k}^{4}\nn\\
&\!-\!24 {k}^{2} ) {\sigma}^{6}\!-\!(24 {\omega}^{7}\!-\!132 {\omega}^{6}\!+\!8(9 {k}^{2}\!+\!26) {\omega}^{5}\!-\!4(63 {k}^{2}\!+\!1){\omega}^{4}\!+\!8 (9 {k}^{4}\!+\!19 {k}^{2}\!-\!22) {\omega}^{3}\!-\!2(54 {k}^{4}\!-\!31 {k}^{2}\!-\!12) {\omega}^{2}\nn\\
&\!+\!(24{k}^{6}\!-\!56 {k}^{4}\!+\!156 {k}^{2}\!+\!48 ) \omega\!+\!12 {k}^{6} \!+\!270 {k}^{4} \!-\!216 {k}^{2}) {k}^{2}{\sigma}^{4} \!+\!(24{\omega}^{5} \!-\!108 {\omega}^{4} \!+\!4(12 {k}^{2} \!+\!29) {\omega}^{3}\!-\!4(24 {k}^{2} \nn\\
& \!-\!11) {\omega}^{2} \!+\! 8(3{k}^{4} \!-\!2 {k}^{2} \!-\!7)\omega\!+\!12 {k}^{4}\!+\!88 {
k}^{2}\!-\!24){k}^{4}{\sigma}^{2}\!-\!4  (2 \omega\!+\!1)({k}^{2}\!+\!4\!-\!4 \omega\!+\!{\omega}^{2} ) {k}^{6} \,.
\end{aligned}\eea
\section{Functions of $P'_2$ polynomial}
\bea\begin{aligned}
E_0&\!\!=\!\!\big[\!-\!\big(430272 {\sigma}^{12} \!\!+\!\! (232704 {k}^{2} \!-\!  84096){\sigma}^{10} \!\!-\!\!  (70608 {k}^{4} \!-\! 285672 {k}^{2} \!+\! 696405){\sigma}^{8}\!\!-\!\! (2048 {k}^{6}\!\!+\!\! 176160 {k}^{4}\!\!-\!\!  573856 {k}^{2} \nn\\
& \!+\!  158790){\sigma}^{6} \!\!+\!\!  (6144 {k}^{6} \!+\! 129488 {k}^{4} \!-\! 
6392 {k}^{2} \!-\! 32544){\sigma}^{4} \!\!-\!\! (6144 {k}^{6} \!+\! 12928 {k}^{4}\!+\! 65216 {k}^{2}\!-\! 2880){\sigma}^{2}\!-\! 11008 {k}^{4} \nn\\
&  \!+\! 16128 {k}^{2} \!+\!  2048 {k}^{6}\big)\!\Delta \!+\! 5614272 {\sigma}^{14} \!\!+\!\!(3034368 {k}^{2}\!+\! 3341952){\sigma}^{12}\!-\! (923472 {k}^{4} \!-\!  5876136 {k}^{2}\!+\!  13288653){\sigma}^{10}  \nn\\
&\!+\! (\,30720 {k}^{6} \!+\!  1409376 {k}^{4} \!-\!  5291328 {k}^{2}\!-\! 1399494)\,{\sigma}^{8}- (116736 {k}^{6}\!-\!  1164336 {k}^{4}\!+\! 7862728 {k}^{2}\!-\! 3083592)\,{\sigma}^{6}\nn\\
&\!+\! (\,165888 {k}^{6}
\!-\! 3181376 {k}^{4}+ 5537120 {k}^{2}\!-\! 381312)\,{\sigma}^{4} - (104448 {k}^{6} \!-\!  1849600 {k}^{4} +  1422592 {k}^{2}+  18432)\,{\sigma}^{2}\nn\\
& \!-\!129024 {k}^{2} \!-\!  318464 {k}^{4} \!+\! 24576 {k}^{6}\big]/5614272\,,\nn\\
E_1&\!\!=\!\!\big[\! \big(  \!-\!29088 {\sigma}^{10} \!\!+\!\! (17652 {k}^{2}  \!-\!56073){\sigma}^{8}  \!-\! (768 {k}^{4} \!+\!15280 {k}^{2} \!+\!5218){\sigma}^{6}  \!+\! (2048 {k}^{4} \!-\!8484 {k}^{2}\!+\!8000){\sigma}^{4} \!+\! (7584 {k}^{2}  \nn\\
& \!-\! 1792 {k}^{4}  \!-\! 1744){\sigma}^{2}  \!-\! 1472 {k}^{2}  \!+\! 512 {k}^{4}\big)\Delta  \!+\! 379296 {\sigma}^{12}  \!-\! (230868 {k}^{2}  \!-\! 1034721){
\sigma}^{10}\!+\!(11520 {k}^{4}\!-\!3072 {k}^{2}  \nn\\
& \!+\!410238){\sigma}^{8}  \!+\!  (478852 {k}^{2} \!-\!
255560 \!-\!35840 {k}^{4}){\sigma}^{6}  \!+\!  (43264 {k}^{4}\!+\!328944 {k}^{2} \!+\!5488){\sigma}^{4}  \!+\!  (98112 {k}^{2} \!+\!10816\nn\\
& \!-\!25088 {k}^{4}){\sigma}^{2} \!-\!14080 {k}^{2} \!+\!6144 {k}^{4}\big]/701784\,,\nn\\
E_2&\!=\!\big[\big(4413{\sigma}^{8} \!+\! (3370 \!-\! 384{k}^{2}){\sigma}^{6} \!+\! (896{k}^{2} \!-\! 1952){\sigma}^{4} \!+\! ( \!-\!640{k}^{2} \!+\! 640){\sigma}^{2} \!+\! 128{k}^{2}\big)\Delta\!-\!57717{\sigma}^{10}  \!-\!(89622\nn\\
& \!-\!5760{k}^{2}){\sigma}^{8} \!-\!(13952{k}^{2} \!-\!27272){\sigma}^{6}
\!+\!(1472 \!+\!12160{k}^{2}){\sigma}^{4} \!+\!( \!-\!3328 \!-\!5504{k}^{2}){\sigma}^{2} \!+\!1536{k}^{2}\big]/350892\,,\nn\\
E_3&\!=\!32(15{\sigma}^{2}\!+\!4\!-\!\Delta){\sigma}^{2}(\sigma^2-1)^{2}/87723\,,
\end{aligned}
\eea
where $\Delta=\sqrt{171\,{\sigma}^{4}+264\,{\sigma}^{2}-80}$.
\section{About the $h_{\mu\varphi}=0$ gauge in $AdS_3$}
To show that $h_{\mu\varphi}=0$ gauge can be attained by a proper diffeomorphism we use the same method for $WAdS_3$ presented in appendix B of \cite{Anninos:2009zi}. If we write the background metric as
\be
g_{MN}= \left(\begin{array}{cc} g_{\mu\nu} & 0 \\ 0 & \phi(r)  \end{array} \right)\,,\qquad ds_2^2=l^2\big[-(1+r^2)d\tau^2+\frac{dr^2}{1+r^2}\big]\,,\qquad \phi(r)=l^2 r^2\,,
\ee
then the non-zero Christoffel symbols will be
\be
^{(3)}\!\Gamma^\rho_{\mu\nu}=\,^{(2)}\!\Gamma^\rho_{\mu\nu}\,,\qquad ^{(3)}\!\Gamma^\varphi_{\varphi\nu}=\frac12\,\partial_\nu\log\phi(r)\,,\quad ^{(3)}\!\Gamma^\nu_{\varphi\varphi}=-\frac12\,\phi(r)\,\partial^\nu\log\phi(r)\,.
\ee
Consider the Fourier expansion of small perturbations in $\varphi$ direction  as follows
\be \label{pert}
h_{\mu\nu}=\int dk\, h^{(k)}_{\mu\nu}\, \exp(ik\varphi)\,, \qquad
h_{\mu\varphi}=\int dk\, h^{(k)}_{\mu\varphi}\, \exp(ik\varphi)\,, \qquad
h_{\varphi\varphi}=\int dk\, h^{(k)}_{\varphi\varphi}\, \exp(ik\varphi)\,.
\ee
The gauge transformation in the linearized theory is given by
\be \label{diffh} \delta_{\zeta}\,h_{MN}=\nabla_M\zeta_N+\nabla_N\zeta_M\,,\ee
which represents the change of the metric perturbation under an infinitesimal diffeomorphism along the vector field $\zeta^{M}$. In fact the right hand side of (\ref{diffh}) is the Lie derivative of the background metric along the vector field $\zeta^{M}$,\ie ${\cal L_{\zeta}}g_{MN}=2\nabla_{(M}\zeta_{N)}$. So under a general diffeomorphism 
\be
\zeta_M=\int dk\, \zeta^{(k)}_M\, \exp(ik\varphi)\,,
\ee
the variation of Fourier modes will be
\be \label{delta}
\delta h^{(k)}_{\mu\nu}=\nabla_\mu\zeta^{(k)}_\nu+\nabla_\nu\zeta^{(k)}_\mu\,,\quad
\delta h^{(k)}_{\mu\varphi}=\partial_\mu \zeta^{(k)}_\varphi+ik \zeta^{(k)}_\mu-\partial_\mu\log\phi(r) \zeta^{(k)}_\varphi\,,\quad
\delta h^{(k)}_{\varphi\varphi}=2ik \zeta^{(k)}_\varphi+\partial^\mu\phi(r)\,\zeta^{(k)}_\mu\,.
\ee

In order to fix the gauge to $h_{M\varphi}=0$ for all values of $k$, we can fix the diffeomorphism modes from two last equations such that 
\be h^{(k)}_{\mu\varphi}=h^{(k)}_{\varphi\varphi}=0\,.\ee

Note that unlike the TMG, from the last equation in (\ref{delta}) we see that for $k=0$ we can fix the
diffeomorphism modes again. 

\section{Frequency conditions}
We have used a change of variable as $\omega=\pm u^\frac12+\frac12$ but since the energy condition restricts us to $Re(\omega)>\frac12$ so only the plus sign is allowed and the reality condition for the frequencies translates to $Re(u)>0$.

\subsection{$P_2$ polynomial (Massive modes)}
Although the $P_2$ polynomial describes the negative energy modes but in what follows we will check the regions of real frequencies for massive modes. The roots of  $P_2$ are given by
\bea
&&u_\pm=\frac{\Delta_1\pm4{\Delta_2}^\frac12}{4\sigma^2(11\sigma^2-2)}\,,\qquad
\Delta_1=\sigma^2(96\sigma^4+31\sigma^2-10)-4(\sigma^2-1)(11\sigma^2-1)k^2\,,\nn\\
&&\Delta_2=\sigma^4(4-68\sigma^2+409\sigma^4-1116\sigma^6+1500\sigma^8)+2\sigma^2(9\sigma^2-2)(\sigma^2-1)^2k^2+(\sigma^2-1)^2k^4\,.
\eea
As we mentioned before, to have a real value for $\omega$, we must have positive real values for $u_\pm$. According to the roots of $\Delta_2$ we have two situations:
\subsubsection*{${\bf{\bullet}}$ When $\Delta_2$ has two real roots for $k^2$ or $0\leq\sigma\leq \sqrt{\frac{2}{11}}$}

1. To have a real valued $u_\pm$ one needs $\Delta_2\ge0$. Suppose that solving $\Delta_2=0$ in terms of $k^2$ gives two real roots, therefore we can  write $\Delta_2=(\sigma^2-1)^2 (k^2-\Sigma_-)(k^2-\Sigma_+)$. So the reality condition of $u_\pm$ will restrict $k^2$ to (note that $\Sigma_->\Sigma_+$)
\bea\label{sigmapm}
&&S_1=\big\{0\leq k^2 \leq \Sigma_{+}\big\}\,,\quad\qquad S_2=\big\{k^2\ge \Sigma_{-}\big\}\,,\nn\\
&&\Sigma_\pm=\sigma^2\frac{(\sigma^2-1)(2-9\sigma^2)\pm\sigma \big(3(2-11\sigma^2)(43\sigma^4-20\sigma^2+4)\big)^\frac12}{\sigma^2-1}\,.
\eea

2. On the other hand $\Delta_2$ itself must be real, which means that $\Sigma_\pm$ must be real. This will be possible if we choose $0\leq\sigma\leq \sqrt{\frac{2}{11}}$ in the admissible interval of $0\leq\sigma<2$.

3. We must notice that in the interval  $0\leq\sigma\leq \sqrt{\frac{2}{11}}\cong 0.426$ we always have $\Sigma_-\geq 0$, but for $0.381\leq\sigma \leq 0.398$ we find that $\Sigma_+<0$. So in this sub-interval just  $k^2\ge \Sigma_{-}$ is acceptable.

4. By looking to the values of $u_\pm$, in the interval $0\leq\sigma\leq \sqrt{\frac{2}{11}}$ we see that the denominator has a negative value for both $u_\pm$. Therefore $u_\pm\geq 0$ if $\Delta_1\pm 4\Delta_2^\frac12\leq0$. In this interval $\Delta_1$ changes its sign from negative values to positive values and it is a monotonically increasing function for all values of $k^2$, and we also have $-4k^2\leq\Delta_1$. Moreover, $4\Delta_2^\frac12$ is a positive monotonically decreasing function for all values of $k^2$ and $ 4\Delta_2^\frac12\leq4k^2$. 

5. Finally we observe that $|4\Delta_2^\frac12|>|\Delta_1|$ for all values in the mentioned interval and for all $k^2$, so we conclude that $u_-$ is the only acceptable solution.

\subsubsection*{${\bf{\ast}}$ Summary:} For $S_1=\big\{0\leq k^2 \leq \Sigma_{+}\big\}$ and $S_2=\big\{k^2\ge \Sigma_{-}\big\}\,,$ with $\Sigma_\pm$ in (\ref{sigmapm}) we have the following table
\begin{table}[ht]\renewcommand{\arraystretch}{1.5}
\center
\begin{tabular}{|c|c|c|c|}\hline
 & $0\leq \sigma \leq 0.381$ & $0.381\leq \sigma \leq 0.398$ & $0.398\leq \sigma \leq 0.426$ \\ \hline
$k^2\in$ & $S_1\cup S_2$ & $S_2$ & $S_1\cup S_2$  \\ \hline
$Solution$ & $u_-$ & $u_-$ & $u_-$  \\ \hline
\end{tabular}
\caption{Summary results for $0\leq\sigma\leq\sqrt{\frac{2}{11}}$.}
\end{table}
\subsubsection*{${\bf{\bullet}}$ When $\Delta_2$ has not any real root for $k^2$ or $\sqrt{\frac{2}{11}}<\sigma<2$}

1. In this case we are restricted to $\sqrt{\frac{2}{11}}<\sigma<2$ where  $\Delta_2$ has not any real root and we find that $\Delta_2> 0$.
Hence for $u_\pm\geq 0$ one needs $\Delta_1\pm 4\Delta_2^\frac12\geq0$.

2. In equation $\Delta_1=\sigma^2(96\sigma^4+31\sigma^2-10)-4(\sigma^2-1)(11\sigma^2-1)k^2$, the first term (coefficient of $k^0$) is always positive in $\sqrt{\frac{2}{11}}<\sigma<2$ but the coefficient of $k^2$ changes its sign from positive values to negative values at $\sigma=1$, so $\Delta_1$ can be either positive or negative.

3. If $|\Delta_1|<4\Delta_2^\frac12$ or equivalently $\Delta_1^2-16\Delta_2<0$ then $u_+$ will be the only allowed solution. Furthermore if $|\Delta_1|>4\Delta_2^\frac12$ or equivalently $\Delta_1^2-16\Delta_2>0$ then we will have two choices. For $\Delta_1>0$ both $u_\pm$ are valid and for $\Delta_1<0$ neither $u_+$ nor $u_-$ are acceptable. 

4. Now consider $\Delta_1^2-16\Delta_2>0$. For $\sqrt{\frac{2}{11}}<\sigma<1$ we always have $\Delta_1>0$ and therefore
\be
\Xi_0<k^2\,,\quad \Xi_0=\frac{\sigma^2 (96\sigma^4+31\sigma^2-10)}{4(\sigma^2-1)(11\sigma^2-1)}\,.
\ee
However, in this interval $\Xi_0<0$ and therefore there is no restriction here. However
\be
\Delta_1^2-16\Delta_2=\sigma^2(11\sigma^2-2)\big(-3\sigma^2(448\sigma^6-640\sigma^4+111\sigma^2+6)-8(\sigma^2-1)(96\sigma^4+43\sigma^2-9)k^2+176(\sigma^2-1)^2k^4\big)\,,
\ee
which by $\Delta_1^2-16\Delta_2>0$ assumption, $k^2$ restricts to
\be\label{xipm}
k^2<\Xi_+\,,\quad k^2>\Xi_-\,,\quad\Xi_\pm=\frac{1}{44(\sigma^2-1)}\big(96\sigma^4+43\sigma^2-9\pm
(81-576\sigma^2+3784\sigma^4+24000\sigma^8-12864\sigma^6)^\frac12\big)\,.
\ee 
In this interval we find that $\Xi_+<\Xi_0<\Xi_-$ therefore $\Xi_+<0$ and consequently $k^2<\Xi_+$ region is not allowed. If we look at $\Xi_-$ we will see that it changes its sign from positive values to negative values at $\sigma=0.506$. The final results are as follows:
\bea
\sqrt{\frac{2}{11}}<\sigma<0.506 \rightarrow \Xi_->0 \rightarrow k^2>\Xi_-\,,\qquad 0.506<\sigma<1 \rightarrow \Xi_-<0 \rightarrow k^2>0\,,
\eea
and two solutions $u_\pm$ are valid.

5. When $1<\sigma<2$ and $\Delta_1>0$ then $\Xi_0>k^2$. In this interval of $\sigma$ we always have  $\Xi_-<\Xi_0<\Xi_+$. Here $\Xi_0>0$ and $\Xi_+>0$ but $\Xi_-$ changes its sign from positive values to negative values at $\sigma=1.103$. Therefore  $u_\pm$ are two allowed solutions, either when $k^2>\Xi_+$ in all the interval or if  $k^2<\Xi_-$ when $1<\sigma<1.103$.

6. Consider $\Delta_1^2-16\Delta_2<0$ where just $u_+$ was valid. For $\sqrt{\frac{2}{11}}<\sigma<1$ we find that $\Xi_+<k^2<\Xi_-$ but since in this interval $\Xi_+<0$ and $\Xi_-$ changes its sign we conclude that for $\sqrt{\frac{2}{11}}<\sigma<0.506$ we have $0<k^2<\Xi_-$. For $1<\sigma<2$ we have $\Xi_-<k^2<\Xi_+$, moreover $\Xi_+>0$ and $\Xi_-$ has a sign change. Particularly for $1<\sigma<1.103$ we have $\Xi_-<k^2<\Xi_+$ and for $1.103<\sigma<2$ we have $0<k^2<\Xi_+$.
\subsubsection*{${\bf{\ast}}$ Summary:} For $S_3=\big\{0\leq k^2\leq\Xi_-\big\}$, $S_4=\big\{\Xi_-\leq k^2\leq\Xi_+\big\}$ and $S_5=\big\{\Xi_+\leq k^2\big\}$ with $\Xi_\pm$ in (\ref{xipm}) we have
\begin{table}[ht]\renewcommand{\arraystretch}{1.5}
\center
\begin{tabular}{|c|c|c|c|c|}\hline
 & $0.426<\sigma\leq 0.506$ & $0.506<\sigma\leq 1$ & $1<\sigma\leq 1.103$ & $1.103<\sigma<2$\\ \hline
$u_+$ solution, $k^2\in$ & $S_3\cup S_4\cup S_5$ & $S_3\cup S_4\cup S_5$ & $S_3\cup S_4\cup S_5$ & $S_3\cup S_4\cup S_5$ \\ \hline
$u_-$ solution, $k^2\in$ & $S_4\cup S_5$ & $S_3\cup S_4\cup S_5$ & $S_3$ & $ S_5$ \\ \hline
\end{tabular}
\caption{Summary results for $\sqrt{\frac{2}{11}}<\sigma<2$.}
\end{table}
\subsection{$P_1$ polynomial (Massless modes)}
The $P_1$ polynomial can be written as $a\, u^3+b\, u^2+c\, u+d=0$ where
\bea
&&a=\sigma^6\,,\quad b=\sigma^4\big(3(\sigma^2-1)k^2+\frac14\sigma^2(2\sigma^2-15)\big)\,,\nn\\
&&c=\sigma^2\big(3(\sigma^2-1)^2k^4+\frac12\sigma^2(\sigma^2-1)(2\sigma^2-17)k^2-\frac{1}{16}\sigma^4(20\sigma^2-59)\big)\,,\nn\\
&&d=(\sigma^2-1)^3k^6+\frac14 (\sigma^2-1)^2 (2 \sigma^2-19) \sigma^2 k^4-\frac{1}{16}\sigma^4(\sigma^2-1)(68\sigma^2-95)k^2+\frac{9}{64}\sigma^6(2\sigma^2-5)\,.
\eea  
As we see in these coefficients, for $0\leq\sigma\leq1$ we  always have $a>0$, $b<0$, $c>0$ and $d<0$. According to the Descartes' rule of signs, this polynomial has either three or one real positive root. So at least there is one real positive root in $0\leq\sigma\leq1$ for all values of $k^2$.

By dividing $P_1$ by $\sigma^6$ and changing  $u$  as $u=t+\frac54-\frac{\sigma^2}{6}-\frac{\sigma^2-1}{\sigma^2} k^2$ we can write the polynomial in its depressed form, $t^3+p\, t+q=0$, i.e.
\be
P_1=t^3-\frac{\sigma^2(\sigma^4+12)+12(\sigma^2-1) k^2}{12\sigma^2}t-\frac{-\sigma^4(\sigma^4-36)+18(\sigma^2-1)(17\sigma^2-6)k^2}{108\sigma^2}=0\,.
\ee
The roots of this equation depend on the sign of its discriminant $\Delta=\frac{q^2}{4}+\frac{p^3}{27}$. If $\Delta>0$ then there will be just one real root, $t=(-\frac{q}{2}+\Delta^\frac12)^\frac13+(-\frac{q}{2}-\Delta^\frac12)^\frac13$ and if $\Delta\leq0$ then there are three real roots for $t$. The discriminant is
\be
\Delta=-\frac{(\sigma^4-4)^2}{432}-\frac{(\sigma^2-1)(3\sigma^8-\sigma^6-98\sigma^4+36\sigma^2+24)}{216\sigma^2}k^2+\frac{(863\sigma^4-612\sigma^2+60)(\sigma^2-1)^2}{432\sigma^4}k^4-\frac{(\sigma^2-1)^3}{27\sigma^6}k^6\,.
\ee
It is very hard to find exactly where we have a sign change  in $\Delta$ but since $\sigma$ is limited between 0 and 2 we can compare values of $k$ with $\sigma$ in three regions.

\subsubsection*{$\bullet\, k\rightarrow 0$}
In this case, only the first term in $\Delta$ is dominant ($\Delta\cong -\frac{(\sigma^4-4)^2}{432}$) so for all values of $\sigma$ its value is negative and we have three roots for $t$. In this region $u\cong t+\frac54+\frac{\sigma^2}{6}$ and $t$ has three positive roots $t\sim\frac14\,,\frac94\,,\frac54-\frac12\sigma^2$.
\subsubsection*{$\bullet\, k\approx O(\sigma)$}
In this region there are sub-regions where we have positive or negative values of $\Delta$. For example if we consider $k=\sigma$ then $\Delta=\frac14-\frac74\sigma^2+\frac{215}{48}\sigma^4-5\sigma^6+\frac{145}{72}\sigma^8-\frac{1}{72}\sigma^{10}$ where we have two roots at $\sigma=0.60$ and $\sigma=1.05$, between these roots $\Delta<0$ and beyond that it is positive. 
\subsubsection*{$\bullet\, k\gg\sigma$}
At this limit the last term in $\Delta$ is dominant and we have $\Delta\cong -\frac{(\sigma^2-1)^3}{27\sigma^6}k^6$, so we have a sign change around the $\sigma=1$. For $\sigma<1$ there is one real positive root and for $\sigma>1$ there are three negative real roots. All roots are close the value of $u=-k^2\frac{\sigma^2-1}{\sigma^2}$.

For clarifying behavior of the roots, we have presented some numerical values of $u$ in the following table

\begin{table}[ht]\renewcommand{\arraystretch}{1.5}
\center
\begin{tabular}{|c|c|c|c|c|}\hline
$\sigma$ & $\frac13$ & $\frac23$ & $\frac43$ & $\frac53$\\ \hline
$k=0$ & $0.25, 2.25, 1.19$ & $0.25, 2.25, 1.03$ & $0.25, 2.25, 0.36$ & $0.25, -0.14, 2.25$ \\ \hline
$k=1$ & $9.97$ & $2.00$ & $2.26$ & $2.30$ \\ \hline
$k=2$ & $33.9$ & $5.90$ & $1.67$ & $1.39$ \\ \hline
$k=3$ & $73.9$ & $12.1$ & $0.23$ & $-0.83$ \\ \hline
$k=10$ & $802$ & $126$ & $-34.6$ & $-52.8$ \\ \hline
\end{tabular}
\caption{Real roots in different regions of parameters space.}
\end{table}

These numerical values show that for $0\leq\sigma\leq1$ we always have at least one positive root and by increasing $k$ only one positive root remains. But for $1<\sigma<2$ by increasing $k$ the number of positive roots reduces and for large enough  values of $k$ there is no positive root in this region.



\begin{figure}[ht]
	\center
		\includegraphics[height=65mm]{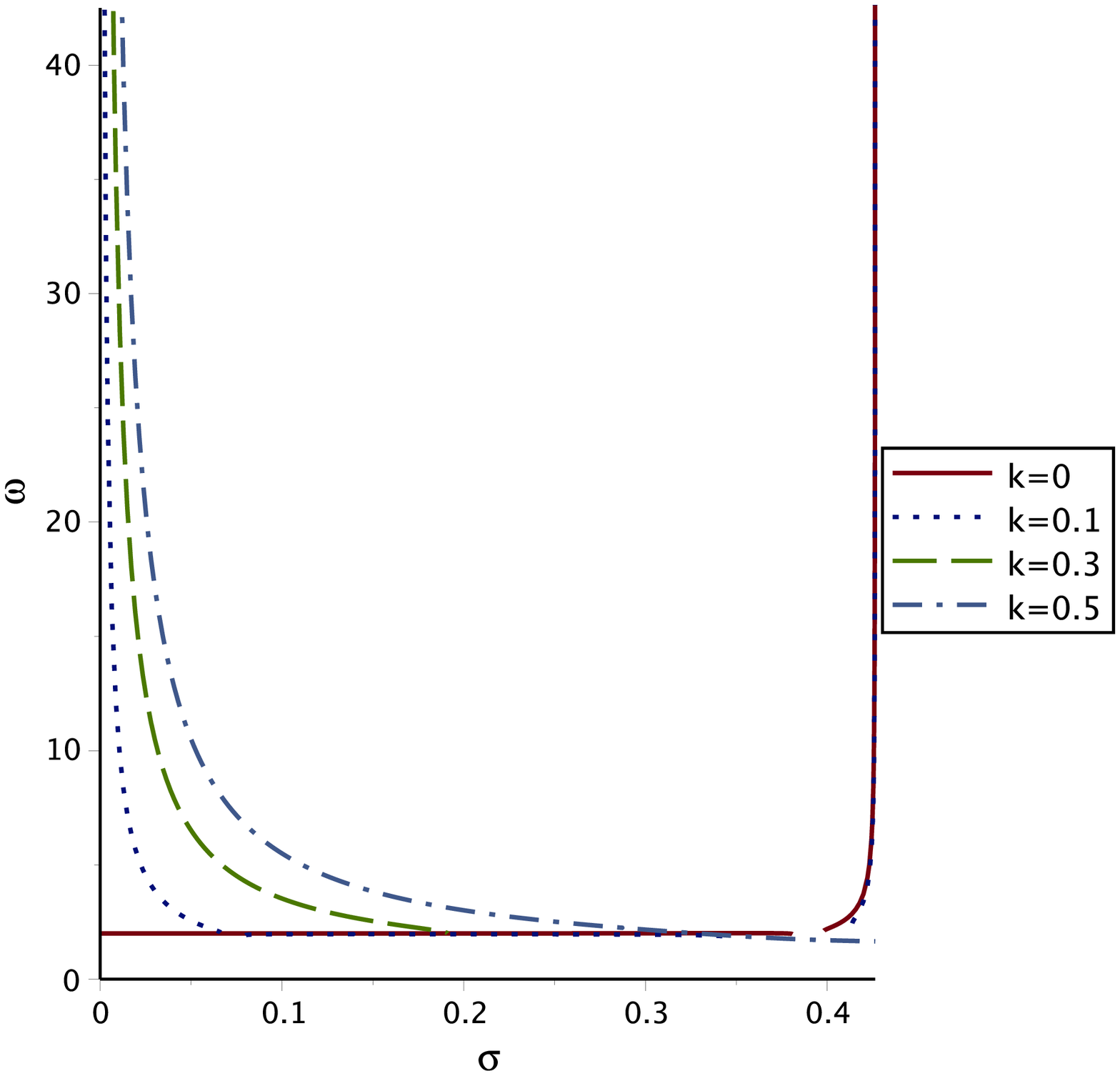}
		\includegraphics[height=65mm]{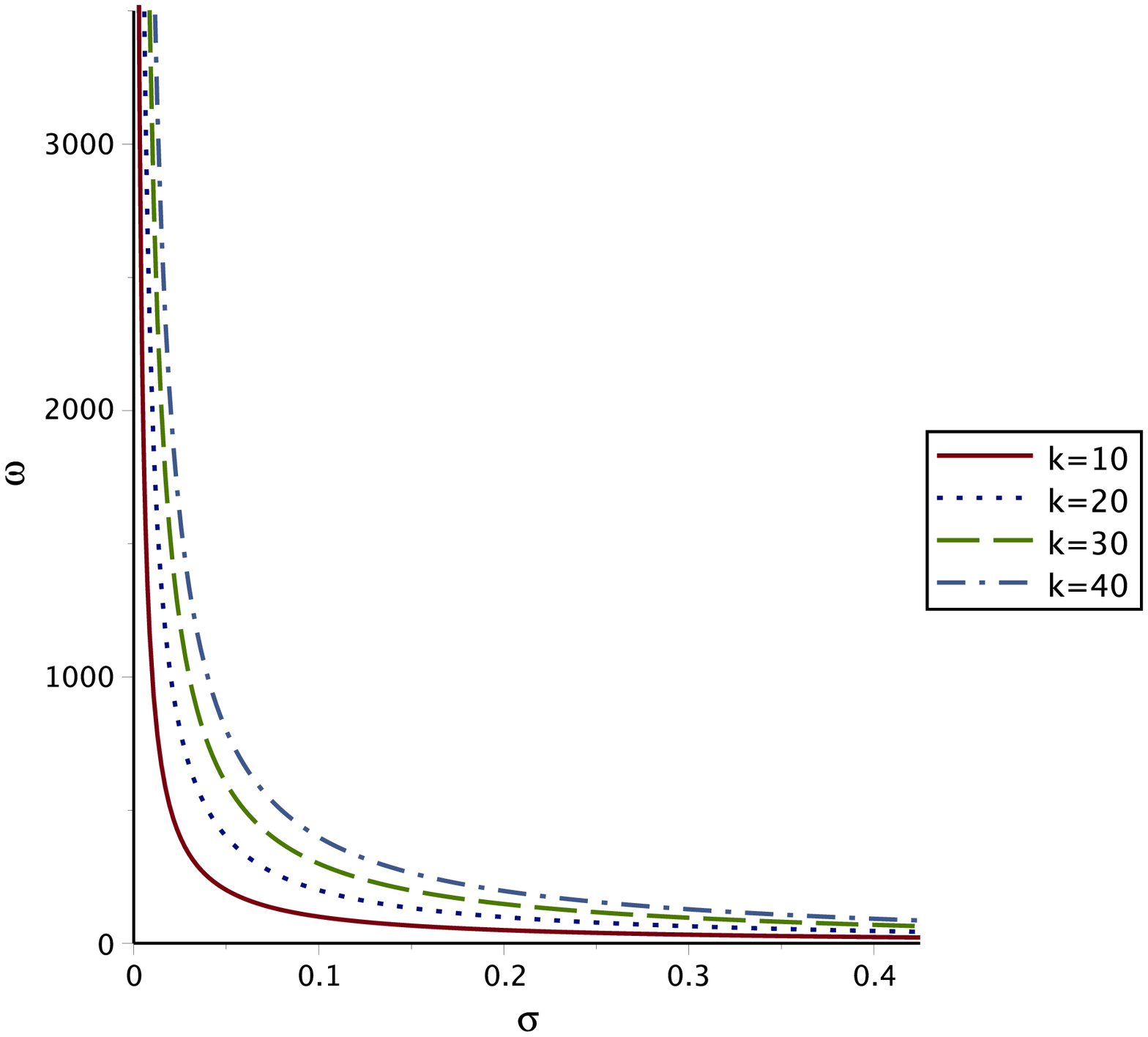}
			\caption{Allowed frequencies of massive modes in $0\leq\sigma\leq\sqrt{\frac{2}{11}}$, according to the table 4.	}
\end{figure}
\begin{figure}[ht]
	\center
		\includegraphics[height=65mm]{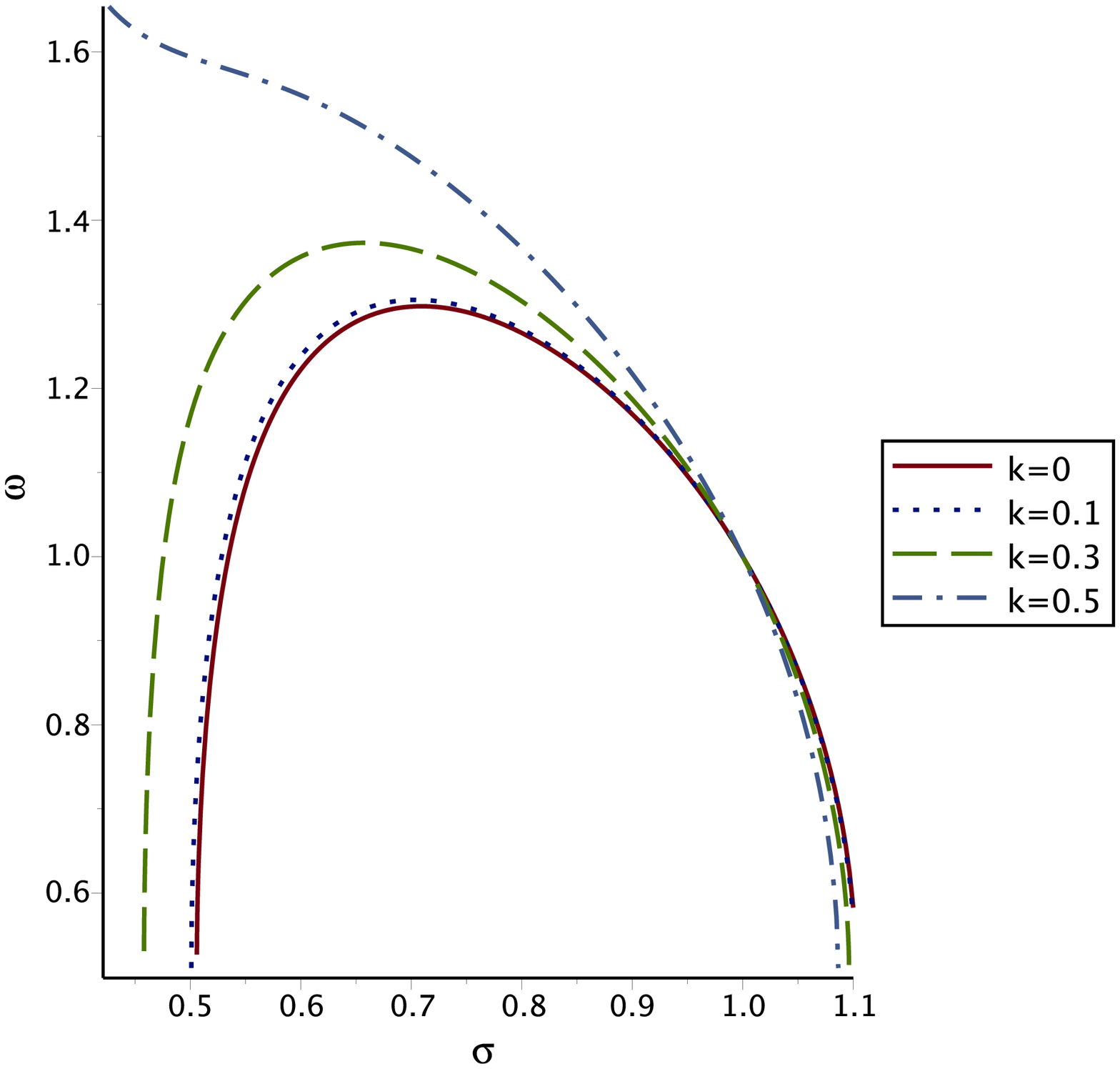}
		\includegraphics[height=65mm]{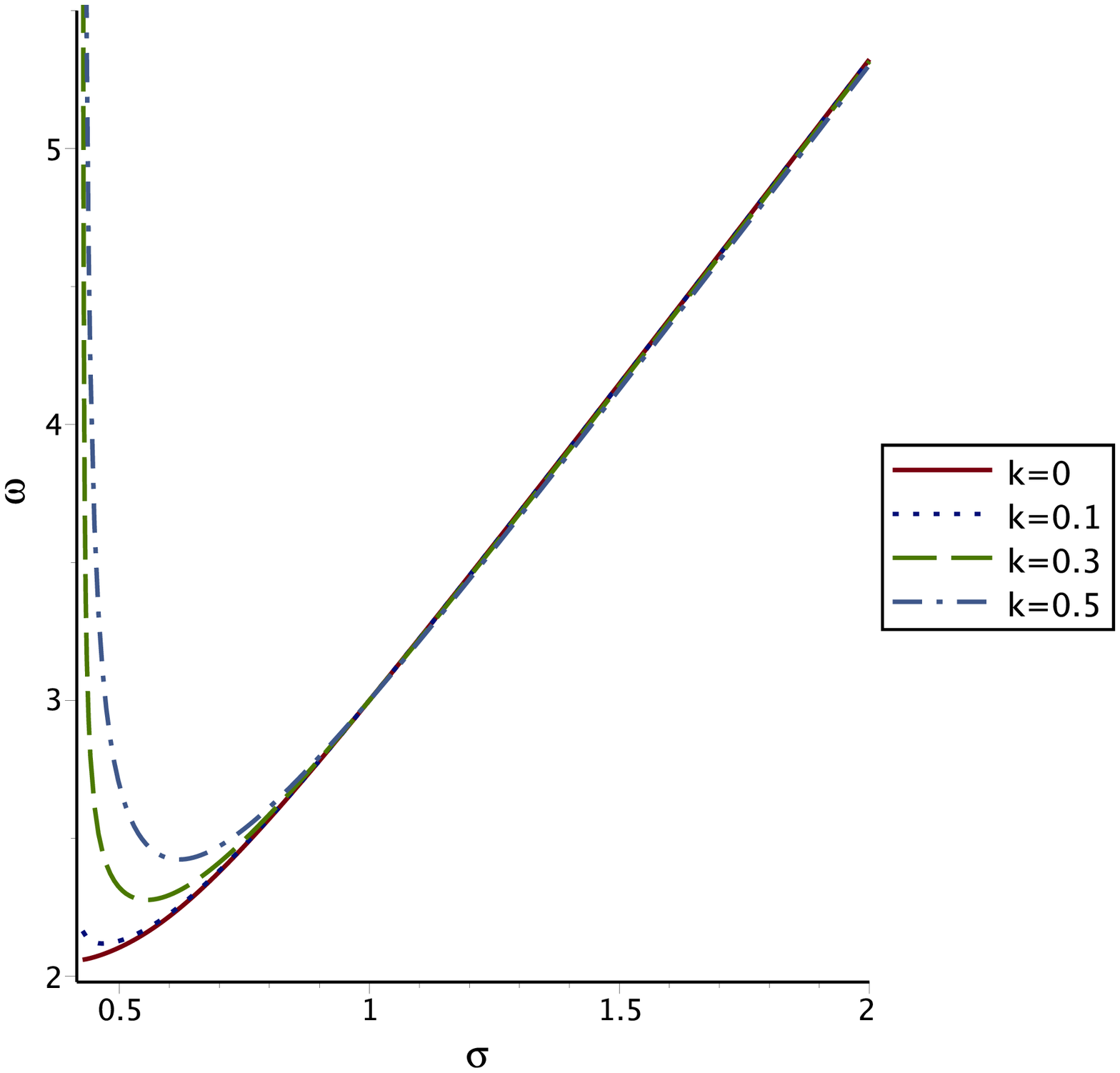}
			\caption{For $\sqrt{\frac{2}{11}}\leq\sigma\leq 1.103$ there are two allowed frequencies. For $1.103\leq\sigma< 2$ only one of them survives (see table 5).}
\end{figure}
\begin{figure}[ht]
	\center
		\includegraphics[height=65mm]{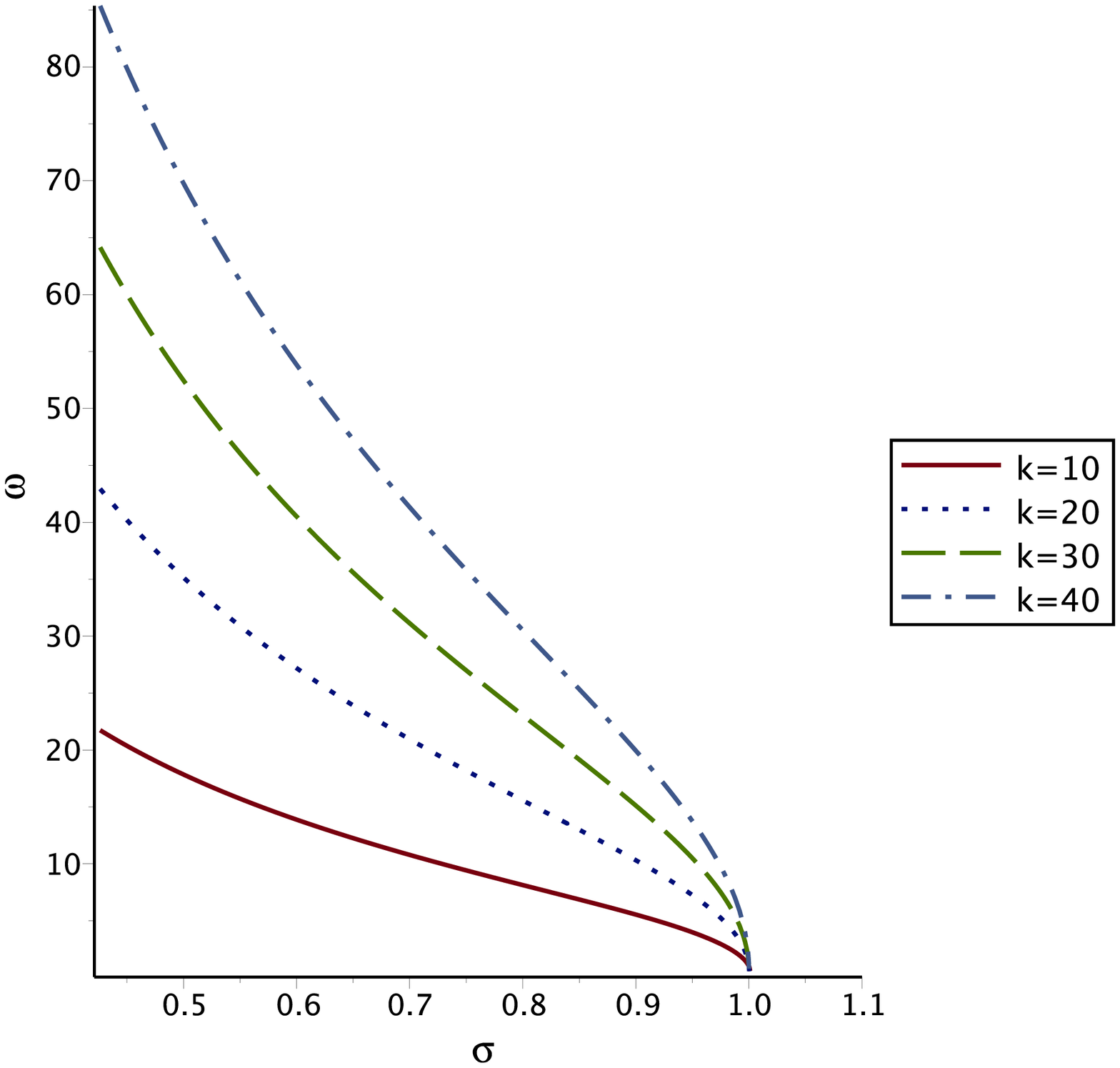}
		\includegraphics[height=65mm]{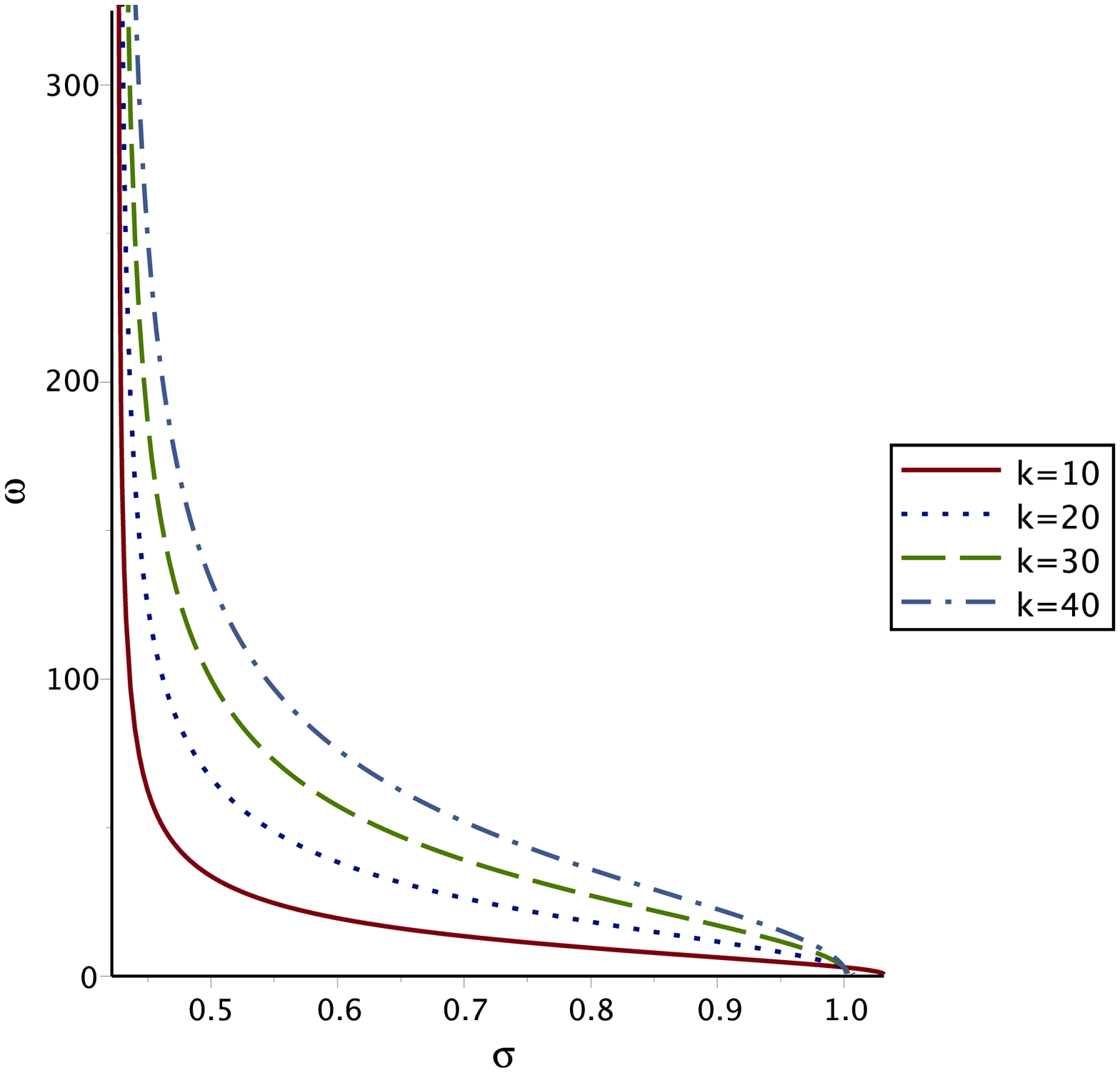}
			\caption{For $\sqrt{\frac{2}{11}}\leq\sigma\leq 1.103$ there are two allowed frequencies.}
\end{figure}

\end{document}